\documentclass[prb,twocolumn,showpacs,amsmath,longbibliography,superscriptaddress,amssymb]{revtex4-1}
\usepackage{graphicx}
\usepackage{dcolumn}
\usepackage{bm}
\usepackage{wasysym}
\usepackage{dsfont}
\usepackage{enumitem}
\usepackage{hyperref}
\usepackage{braket}

\newcommand{\beq}{\begin{equation}}
\newcommand{\eeq}{\end{equation}}
\newcommand{\beqn}{\begin{eqnarray}}
\newcommand{\eeqn}{\end{eqnarray}}

\newcommand{\vphi}{\varphi}
\newcommand{\cA}{ {\cal A} }

\newcommand{\cC}{ {\cal C} }

\newcommand{\cE}{ {\cal E} }

\newcommand{\cP}{ {\cal P} }
\newcommand{\tcP}{\tilde{\cal P}}

\newcommand{\cV}{ {\cal V} }
\newcommand{\cZ}{ {\cal Z} }
\newcommand{\tcZ}{ \tilde{\cal Z} }
\newcommand{\cT}{ {\cal T}}

\newcommand{\ii}{\mathrm{i}}

\newcommand{\SO}{\mathrm{SO}}

\newcommand{\U}{\mathrm{U}}

\newcommand{\tr}{\mathrm{tr}}
\newcommand{\cM}{{\cal M}}
\newcommand{\cEx}{|{\cal E}_X\rangle}
\newcommand{\cI}{\mathcal{I}}
\newcommand{\twotheta}{(\theta_1,\theta_2)}

\usepackage{xcolor}

\definecolor{forestgreen}{rgb}{0.13, 0.55, 0.13}

\begin{document}

\title{Decoding coherent errors in toric codes on honeycomb and square lattices: duality to Majorana monitored dynamics and symmetry classes}

\author{Zhou Yang}
\affiliation{Department of Physics, Cornell University, Ithaca, New York 14853, USA}

\author{Andreas W. W. Ludwig}
\affiliation{Department of Physics, University of California, Santa Barbara, California 93106, USA}

\author{Chao-Ming Jian}
\affiliation{Department of Physics, Cornell University, Ithaca, New York 14853, USA}

\begin{abstract}

Topological stabilizer codes, such as the toric and surface codes, are leading candidates for fault-tolerant quantum computation. While their decodability under stochastic noise has been extensively studied, the effects of coherent errors, which involve quantum interference, remain less explored. In this work, we study the decodability of toric codes on honeycomb and square lattices subject to $X$- and $Z$-type coherent errors generated by the $X$- and $Z$-rotations on each qubit. We establish a duality between these decoding problems and 1+1D monitored dynamics of non-interacting Majorana fermions. This duality shows that the Altland–Zirnbauer symmetry class of the dual Majorana dynamics governs the universal structure of the decodability phase diagram. We show that the honeycomb-lattice toric code (hTC) with $X$-type error is dual to class-DIII dynamics, while the hTC with $Z$-type error and the square-lattice toric code (sTC) with both error types are dual to class-D dynamics. The key distinction arises from time-reversal symmetry. In class DIII, the generic transition out of the decodable phase is dual to a measurement-induced transition between dynamical phases with area-law and logarithmic entanglement scaling. In contrast, in class D, the generic decodability transition corresponds to a transition between two topologically distinct area-law phases. To explore these transitions in microscopic models, we consider hTC and sTC with $X$-type errors as representatives and introduce a minimal two-parameter coherent error model with spatially varying rotation angles. Using analytical and numerical methods, we map out the decodability phase diagrams and characterize the universal behavior of the transitions. We find that the decodability of sTC is more vulnerable to spatially varying coherent errors than uniform ones.

\end{abstract}

\date{\today}

\maketitle

\tableofcontents

\section{Introduction}

Quantum error correction is essential for reliable quantum information processing in noisy devices. Among the most prominent architectures are topological stabilizer codes, such as the toric code and surface code \cite{Kitaev2003}, which encode logical information in nonlocally entangled physical qubits. A central question for these codes is their decodability, which concerns whether physical errors can be corrected to recover the encoded logical information. Extensive work has studied topological codes under incoherent Pauli errors, such as stochastic bit and phase flips. At low error rates, the system remains in a decodable phase where logical information can be reliably recovered, while higher error rates drive the code into undecodable phases characterized by logical failures. The pioneering work of  Ref.~\onlinecite{Dennis2002topo} established that these phases are separated by sharp transitions in the thermodynamic limit. Such decodability transitions can be understood through disordered statistical models describing the distribution of stabilizer violations, i.e., syndromes, induced by errors. More recently, they have also been linked to decoherence-induced phase transitions in the error-corrupted mixed states of noisy topological codes \cite{BaoFanError_Field_Double,FanBaoTopoMemory,LeeJianXu2023}.

Beyond incoherent (stochastic) noise, coherent errors arising from imperfect gate control can also significantly affect the performance of topological codes \cite{Kueng2016,WallmanFlammia2015,Wallman2015,bravyi2018correcting,Venn_2023,BaoUnitaryErrors2024,BehrendsBeri2025,BealeLaflame2018,Gottesman2019MaximalSensitive,WallmanEmerson2016,HuangFlammia2019,BehrendsSurfaceCodesQuantumCircuits,BaoUnitaryErrors2024,Cheng_2025}. Coherent errors can induce interference effects in syndrome distributions, leading to qualitatively different behavior from the stochastic or incoherent cases. How these interference effects influence the decodability of topological stabilizer codes remains not fully understood in general. Nevertheless, the standard Pauli decoding framework for incoherent errors can be naturally extended to the coherent setting: one measures syndromes and applies syndrome-dependent Pauli-string corrections to recover the logical information. This framework is expected to be feasible for realistic quantum devices\cite{GoogleAISurfaceCode_s,GoogleBelowThreshold_s,Bluvstein2024,Iqbal2024MeasurementFeedforward}. It is therefore both timely and valuable to study the decodability of topological codes under coherent errors within this framework. In principle, more general decoding strategies involving non-local and non-Pauli recovery operations could be considered. However, their practical advantages and resource requirements in realistic implementations remain unclear. For this reason, we focus in this work on the standard Pauli decoding framework.

Recently, significant progress has been made in understanding the decoding problem of the two-dimensional square-lattice surface code subject to coherent errors \cite{bravyi2018correcting,Venn_2023,BehrendsBeri2025,BehrendsSurfaceCodesQuantumCircuits,BaoUnitaryErrors2024,Cheng_2025}, particularly for uniform single-qubit coherent errors, where each physical qubit independently undergoes the same unitary rotation.  Ref.~\onlinecite{bravyi2018correcting} analyzed the decodability of this system using specific decoding algorithms. Notably, for coherent errors generated by single-qubit Pauli-$X$ rotations, Refs.~\onlinecite{Venn_2023,BehrendsSurfaceCodesQuantumCircuits} established
a connection between the disordered statistical model governing the code's decodability (within the Pauli decoding framework) and the two-dimensional Anderson localization problem in the Altland–Zirnbauer (AZ) symmetry class D. This correspondence points to a close relationship between decodability transitions and localization transitions.

This observation raises several important questions. How general is the connection between surface codes with coherent errors and the Anderson localization problems? Is there a precise correspondence between the phases in these systems? What determines the AZ symmetry class of a given decoding problem? More broadly, coherent errors need not be spatially uniform. Can spatial inhomogeneity give rise to new structures in the decodability phase diagram?

In this work, we address these questions by studying both the honeycomb-lattice toric code (hTC) and the square-lattice toric code (sTC), which are equivalent to surface codes on the corresponding lattices up to boundary conditions. We consider both $X$- and $Z$-type single-qubit coherent errors generated by Pauli-$X$ and Pauli-$Z$ operators, without assuming spatial uniformity. To analyze the decoding problem, we establish a duality between these toric codes and monitored quantum dynamics of Majorana fermions in 1+1D. Within this dual framework, the syndrome probability distribution is mapped to the distribution of quantum trajectories in the monitored dynamics. Consequently, phases in the decodability diagram correspond to distinct dynamical phases, and decodability transitions are identified with measurement-induced phase transitions in the monitored dynamics.

An important insight from this duality is that the AZ symmetry class of the dual Majorana monitored dynamics \cite{JianRTN2022,bhuiyan2025free} plays a central role in organizing the structure of the decodability phase diagram. Time-reversal (TR) symmetry is a key factor for determining the symmetry classes. Without TR symmetry, the error-corrupted toric code maps to the Majorana monitored dynamics in symmetry class DIII, which preserves only fermion parity. This scenario applies to the hTC subject to $X$-type coherent errors. When the error-corrupted toric code respects the TR symmetry, the dual monitored dynamics also becomes TR-symmetric, placing it in symmetry class D. This scenario applies to the hTC with $Z$-type coherent errors as well as to the sTC with both types of errors. In all cases considered here, the dual Majorana monitored dynamics are non-interacting, so their universal behavior and possible phases are dictated by their AZ symmetry classes.

This duality also clarifies the relation between decoding and Anderson localization problems. As noted in  Ref.~\onlinecite{JianRTN2022}, monitored
dynamics of non-interacting fermions and Anderson localization problems can be unified under the same AZ symmetry classification. However, their universal properties can still differ due to the distinct probability measures governing the quantum trajectories of the monitored dynamics (dual to the syndrome distributions in the toric codes) and the disorder realizations in the Anderson problems. In replica-based analysis, this distinction manifests in different replica limits. In our setting, this difference is particularly important in symmetry class D.

With the symmetry class of the decoding problems established through their dual description in terms of Majorana monitored dynamics, a global perspective of the phase diagram becomes transparent. It is known that class-DIII non-interacting Majorana systems can exhibit two topologically distinct phases with area-law entanglement scaling, as well as a critical phase characterized by logarithmic entanglement scaling \cite{JianRTN2022,FavaNahum2023,jian2023measurement,Pan_2025,kells2023topological,GYZhuLearning2025,Negari_2024}. In the decoding context, we show that one of these area-law phases corresponds to the decodable phase of the toric code, while the other area-law phase and the critical phase correspond to distinct undecodable regimes. Generic phase transitions in this symmetry class occur between an area-law phase and the critical phase.

In contrast, class-D non-interacting monitored dynamics is expected to exhibit a family of $\mathbb{Z}$-classified area-law  phases~\cite{SchnyderRyu2010NJP,XiaoKawabata2024,Pan_2025,bhuiyan2025free}, one of which corresponds to the decodable phase of the error-corrupted toric code. For symmetry class D, we also show that the analog of the critical phase is unstable under renormalization group (RG) flow. Consequently, the decodability transition in symmetry class D is expected to correspond to a direct transition between distinct area-law phases in the dual Majorana monitored dynamics.

We note that Ref.~\onlinecite{Venn_2023} studied the class-D decoding problem for the sTC with uniform coherent error. Numerical signatures suggestive of a critical phase and an associated decodability transition were reported, seemingly at odds with our expectation. However, our analysis indicates that the previously observed numerical signature might be attributed to a large but finite correlation length, which can be estimated using RG arguments. We further support this interpretation with additional numerical results using concrete error models that will be described below.

The duality to the Majorana monitored circuit not only provides a global perspective on the decodability phase diagrams of error-corrupted toric codes in both symmetry classes, but also helps identify the microscopic coherent error models that can drive the decodability transitions. As representative examples of symmetry classes DIII and D, we consider the hTC and the sTC, respectively, both subject to the $X$-type coherent error. To elucidate the global structure of the decodability phase diagrams, we introduce a minimal two-parameter error model for each symmetry class. These models incorporate spatial variations in the $X$-rotation angles of the physical qubits, with the spatially uniform
coherent error included as a special case. We analyze the decodability phase diagram in these models using a combination of analytical and numerical approaches.

For the hTC in symmetry class DIII, our numerical results confirm that the generic decodability transition out of the decodable phase is dual to an area-law–to–critical phase transition in the corresponding Majorana monitored dynamics, as expected. Interestingly, along a special line in the phase diagram, an emergent time-reversal (TR) symmetry appears, effectively changing the system's symmetry class to D. Along this line, the decodability transition instead corresponds to an area-law–to–area-law transition, consistent with our conclusion that class-D Majorana monitored dynamics does not admit a stable critical phase. 

For the sTC with $X$-type coherent errors, which generically belongs to symmetry class D (without any fine-tunig of error angles), the transition out of the decodable phase is likewise dual to an area-law–to–area-law transition. Notably, this decodability transition occurs away from the uniform-error limit. Hence, it has not been seen in previous studies of sTC with coherent errors, which have primarily focused on the uniform case. The inclusion of spatial variations in the error angles in our two-parameter error model is essential for revealing this class-D decodability transition. More generally, this result provides a concrete example in which spatially non-uniform coherent errors become more detrimental to the code's decodability than uniform ones due to interference effects.

While most of our numerical investigations of the decodability phase diagram are carried out using the dual monitored dynamical system, we also study the class-D decodability transition directly from an error-correction perspective in the square-lattice surface code under the two-parameter coherent error model. Our numerical results on the (post-optimal-decoding) logical error rates as functions of the code distances corroborate the class-D decodability transition, which is dual to an area-law–to–area-law transition in the corresponding Majorana-monitored dynamics.

The rest of the paper is organized as follows. Sec.~\ref{sec:hTC} focuses on the hTC under both $X$-type and $Z$-type coherent errors. In Sec.~\ref{sec:hTCX}, we analyze the disordered statistical model and its dual non-interacting Majorana monitored dynamics that capture the code’s decodability. We show that the monitored dynamics belong to symmetry class DIII and discuss the corresponding phase diagram structure. To elucidate the global features of the decoding problem in this class, we introduce a minimal two-parameter coherent error model. Using a combination of analytical and numerical methods, we map out the resulting decodability phase diagram for $X$-type errors. In Sec.~\ref{sec:hTCZ}, we demonstrate that the decoding problem of the hTC with $Z$-type coherent errors preserves TR symmetry and is therefore dual to non-interacting Majorana monitored dynamics in symmetry class D. We discuss the possible phases and the generic structure of the decodability phase diagram using both the topological classification of area-law phases and a continuum nonlinear sigma model description for this symmetry class. Sec.~\ref{sec:sTC} is devoted to the sTC. We show that, for both $X$-type and $Z$-type coherent errors, the decoding problem is dual to Majorana monitored dynamics in symmetry class D. We again introduce a two-parameter coherent error model to reveal the global structure of the decoding problem. Numerical simulations, performed both in the dual monitored dynamics and directly on the square-lattice surface code, are used to map out the decodability phase diagram.

\section{Honeycomb-lattice Toric Code}
\label{sec:hTC}
\subsection{Stabilizers and Coherent Error Models}
\label{sec:Stabilizer_ErrorModel}
\begin{figure}
    \centering
    \includegraphics[width=0.95\linewidth]{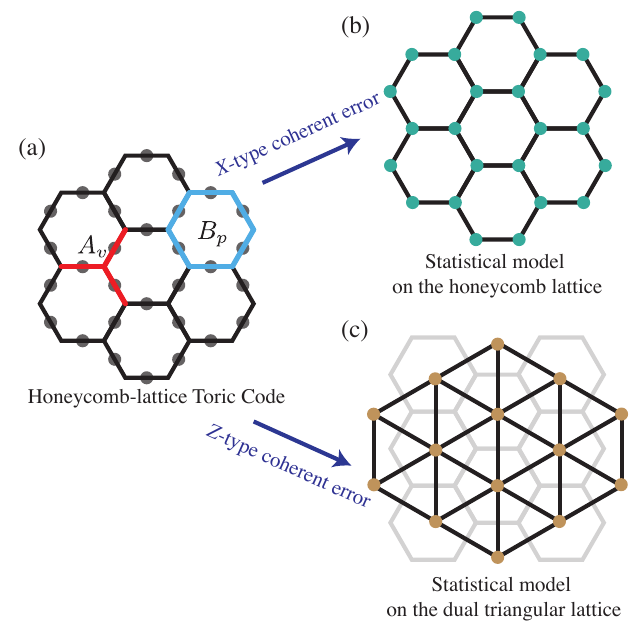}
    \caption{(a) Honeycomb-latice toric code with the qubits (black dots) on the links of the lattice. The vertex stabilizer $A_v$ (red) acts on the three links connected to the vertex $v$. The plaquette stabilizer $B_p$ (blue) acts on the six links of the plaquette $p$. (b) The syndrome distribution generated by the $X$-type coherent error can be calculated from a disordered statistical model with Ising spins (green dots) on the 
    honeycomb lattice. (c) The syndrome distribution generated by the $Z$-type coherent error can be calculated from a disordered statistical model with Ising spins (brown dots) on
    the dual triangular lattice. }
    \label{fig:honeycombTC}
\end{figure}

The honeycomb-lattice toric code (hTC) is a stabilizer code on the honeycomb lattice, with qubits residing on the links. We denote Pauli operators acting on the link $l$ as $X_l$, $ Y_l$, and $Z_l$. The hTC includes a stabilizer  $A_{v}$ for every vertex $v$ and a stabilizer $B_{p}$ for every plaquette of the lattice (as illustrated in Fig. \ref{fig:honeycombTC}(a))
\begin{align}
   A_{v} = \prod_{v\in l } X_l,~~~~~~B_{p} = \prod_{l\in p} Z_l.
\end{align}
Here, $\prod_{v\in l}$ is the product over all three links $l$ connected to the vertex $v$ and $\prod_{l\in p}$ is the product over the six links $l$ on the plaquette $p$. A state $|\Omega\rangle$ in the code space, i.e., a logical state, is a simultaneous eigenstate of all $A_{v}$'s and $B_{p}$'s with eigenvalues $+1$:
\begin{align}
    A_{v} |\Omega\rangle = B_{p}|\Omega\rangle = |\Omega\rangle,~~~~\forall v, p.
\end{align}
For a state corrupted by errors, the measurements of all the $A_{v}$'s and $B_{p}$'s yield a collection of eigenvalues $\{s_v = \pm 1\}$ and $\{s_p = \pm 1\}$, which are referred to as syndromes. The trivial syndrome is the case where $s_v = s_p = 1$ for all vertices $v$ and plaquettes $p$.

The class of errors we focus on in this work is the single-qubit coherent errors caused by unitary rotations $\prod_l e^{\ii \theta_l X_l}$ or $\prod_l e^{\ii \theta_l Z_l}$, namely every qubit is unitarily rotated about the $X$- or $Z$-axis by an angle $\theta_l$. We refer to the former as the $X$-type coherent error and the latter as the $Z$-type. Under such coherent errors, the error-corrupted states $|{\cal E}_X\rangle$ and $|{\cal E}_Z\rangle$ remain to be coherent pure states:
\begin{align}
    |{\cal E}_X\rangle = \prod_l e^{\ii \theta_l X_l} |\Omega \rangle,~~~~
    |{\cal E}_Z\rangle = \prod_l e^{\ii \theta_l Z_l} |\Omega \rangle.
    \label{eq:ErrorModel}
\end{align}
When the stabilizers are measured, $|{\cal E}_X\rangle$ only exhibts non-trivial syndromes $\{s_p\}$ on the plaquettes, while $|{\cal E}_Z\rangle$ only exhibts non-trivial syndromes $\{s_v\}$ on the vertices. Since the vertex syndromes and plaquette syndromes are often decoded separately in the toric code, the error models in Eq. \eqref{eq:ErrorModel} offer the simplest settings for investigating how coherent errors affect the universal behavior of the code. The angles $\{\theta_l\}$ parameterize the strengths of the errors. $\theta_l$ on each qubit can be independent from each other in principle. The uniform version of such a coherent error, namely $\prod_l e^{\ii \theta X_l}$ or $\prod_l e^{\ii \theta Z_l}$ with a spatially uniform angle $\theta_l = \theta$, was first studied in the square-lattice surface code in Ref. \onlinecite{bravyi2018correcting}. Here, in our study, we will first keep $\{\theta_l\}$ general, and later focus on a two-parameter family of $\{\theta_l\}$ that suffices to capture the universal behavior in terms of the code's decodability.

A general procedure for decoding the toric code (within the Pauli framework) is as follows. First, the syndromes $\{s_p\}$ or $\{s_v\}$ are measured. Then, a decoder uses a certain algorithm to find a syndrome-dependent Pauli-string operation, i.e., a Pauli correction, to error-correct the post-measurement state $\prod_p \left(\frac{1+s_p B_p}{2}\right) |{\cal E}_X\rangle$ or $\prod_v \left(\frac{1+s_v A_v}{2}\right) |{\cal E}_Z\rangle$ with the goal of recovering the initial logical state $|\Omega\rangle$. In general, whether the logical error rate vanishes after decoding depends on the details of the decoder, the code distance (or system size), and the error angles. At large system sizes, one can define a fundamental limit of decodability regarding whether there exists a decoder that can achieve a vanishing logical error rate. This limit, also known as the error threshold, is an intrinsic property of the code and the error model. Ref. \onlinecite {Dennis2002topo} pointed out that an optimal decoder that can achieve a vanishing logical error rate up to the error threshold is the so-called maximum-likelihood decoder. The performance of this decoder and the code's fundamental decodability are tied to the probability distribution of the syndromes. The error threshold manifests as a phase transition out of the decodable phase in this syndrome probability distribution. In undecodable phases, any decoding would result in a finite logical error rate.

In the error-corrupted states $|{\cal E}_X\rangle$ and $|{\cal E}_Z\rangle$, the syndrome probability distribution is respetively given by
\begin{align}
    {\cal P}_X(\{s_p\}) = \langle{\cal E}_X |\prod_p\left(\frac{1+s_p B_p}{2}\right) |{\cal E}_X\rangle, \nonumber  \\
    {\cal P}_Z(\{s_v\}) = \langle{\cal E}_Z |\prod_v \left(\frac{1+s_v A_v}{2}\right) |{\cal E}_Z\rangle.
    \label{eq:SyndromeDist}
\end{align}
Later, we will see that these distributions can be directly expressed in terms of the partition functions of a class of disordered classical statistical models, which can be further dualized to 1+1D monitored dynamics of Majorana fermions. We will study the phases and phase transitions in these statistical models and their dual dynamical systems to provide insights into the decodability phase diagram of the hTC with coherent errors. In particular, we will highlight the role of the AZ symmetry classification of the dual monitored dynamics in controlling the structure of the phase diagram. This phase diagram concerns the universal behavior of error-corrupted codes in the large-system-size limit. It is expected to be independent of the specific choice of the logical state $|\Omega\rangle$ and the code's spatial topology (e.g., torus v.s. disk).

Before discussing the disordered statistical models and their dual monitored dynamics, we introduce an anti-unitary time-reversal (TR) symmetry $\mathcal{T}$ of the hTC:
\begin{align}
    \mathcal{T} \equiv \left(\prod_l X_l\right) \,{\cal K}
    \label{eq:TR}
\end{align}
where $\cal K$ is the complex conjugate operation. This TR symmetry is important for our later discussion of the symmetry classification of the minotored dynamics. Under $\mathcal{T}$, the qubit operators transform as ${\cal T}: X_l\rightarrow X_l, Z_l\rightarrow -Z_l, Y_l\rightarrow Y_l$. All the stabilizers are invariant under $\cal T$, while the coherent errors transform as
\begin{align}
    {\cal T}:~~ 
        & \prod_l e^{\ii \theta_l X_l} \rightarrow \prod_l e^{-\ii \theta_l X_l},~~~~\prod_l e^{\ii \theta_l Z_l} \rightarrow \prod_l e^{\ii \theta_l Z_l}.
        \label{eq:TR_Error}
\end{align}
Notice that this TR symmetry is preserved by the $Z$-type coherent error but not the $X$-type.
Later, we will see that such a difference yields distinct symmetry classes for the Majorana monitored dynamics dual to the decoding problem of $|{\cal E}_{X/Z}\rangle$.

\subsection{$X$-type coherent error}
\label{sec:hTCX}
In the following, we focus on the decoding problem of the hTC subject to $X$-type coherent errors $\prod_l e^{\ii \theta_l X_l}$. We introduce a two-dimensional disordered statistical model and its dual description in terms of 1+1D Majorana monitored dynamics, which together capture the syndrome distribution ${\cal P}_X({s_p})$ and the associated decodability phase diagram. We demonstrate that the symmetry classification of the dual monitored dynamics dictates the general structure of this phase diagram. To make this connection concrete, we consider a minimal two-parameter family of $X$-type coherent errors and analyze it using both analytical and numerical methods. Within this model, we show that a variety of decodability transitions can be realized.

\subsubsection{Statistical Model and Majorana Monitored Circuit}
The syndrome distribution ${\cal P}_X(\{s_p\})$ in $|{\cal E}_X\rangle$ (see Eq. \eqref{eq:SyndromeDist}) can be written in terms of the partition function ${\cal Z}_{X,\{s_p\}}$ of a classical disordered statistical model of classical Ising spins $\tau_v =\pm 1$ living on the vertices on the honeycomb lattice  (see Fig. \ref{fig:honeycombTC}(b)):
\begin{equation}
     {\cal P}_X(\{s_p\}) = |{\cal Z}_{X,\{s_p\}}|^2. \label{eq:PtoZ}
\end{equation}
with the partition function ${\cal Z}_{X,\{s_p\}}$ given by
\begin{align}
    {\cal Z}_{X,\{s_p\}} = {\cal N} \sum_{\{\tau_v =\pm1\}} \exp{\left[ - \sum_{\langle v,v'\rangle} J_{v,v'} \eta_{v,v'} \tau_v \tau_{v'} \right]}.
    \label{eq:hTCX_DisorderedPartitionFunction}
\end{align}
In this disordered statistical model, there is a complex nearest-neighbor Ising coupling $J_{v,v'}$ satisfying $e^{2J_{v,v'}} = \ii \tan\theta_l$ for every link $l=\langle v, v'\rangle$ connecting the neighboring sites $v$ and $v'$. $\cal N$ is an unimportant normalization constant. The variables $\eta_{v,v'} = \pm 1$ capture the random-bond disorder on each link $l=\langle v,v'\rangle$. They are set by the syndrome $\{s_p\}$ via $s_p = \prod_{l \in p} \eta_l$. One can easily show that the different random-bond configurations $\{\eta_l\}$ associated with the same syndrome $\{s_p\}$ are related to each other by gauge transformations $\tau_v\rightarrow b_v \tau_v,~ J_{v,v'}\rightarrow J_{v,v'}b_vb_{v'}$ with $b_v = \pm 1$ for each vertex. In the following, we will also denote ${\cal Z}_{X,\{s_p\}}$ as ${\cal Z}_{X,\{\eta_l\}}$.
The derivation of this statistical model follows the standard method introduced in Ref. \onlinecite{Dennis2002topo} and is summarized in App. \ref{app:statmodel}.  More recent examples that concern the uniform coherent error on the square-lattice surface code can be found in Refs. \onlinecite{Venn_2023,BehrendsSurfaceCodesQuantumCircuits}.

The decodability of the error-corrupted code needs to be evaluated by averaging over all possible syndrome configurations. From the disordered statistical model perspective, we need to consider the disorder-averaged $F_X$ of the free energy $-\log |{\cal Z}_{X,\{s_p\}}|^2$ weighted by the syndrome probability ${\cal P}_X(\{s_p\})$:
\begin{align}
    F_X = \sum_{\{s_p\}} -{\cal P}_X(\{s_p\}) &\log |{\cal Z}_{X,\{s_p\}}|^2.
\end{align}
Singularities in $F_X$ indicate the phase transition in the decodability of the hTC corrupted by $X$-type coherent errors.  The free energy $F_X$ can be further rewritten as 
\begin{align}
    F_X \propto  &\sum_{\{\eta_l\}} -|{\cal Z}_{X,\{\eta_l\}}|^2\log |{\cal Z}_{X,\{\eta_l\}}|^2.
    \label{eq:hTCX_FreeEnergy}
\end{align}
This relation holds because all the gauge-equivalent random bond configurations $\{\eta_l\}$  yield the same partition function ${\cal Z}_{X,\{\eta_l\}} = {\cal Z}_{X,\{s_p\}}$ with the syndrome $s_p=\prod_{l\in p} \eta_l$. Also, the number of gauge-equivalent random bond configurations is the same across all syndromes. The technical advantage of the right-hand side of this equation is that each $\eta_l$ is treated as an independent variable, which will be helpful for mapping this disordered statistical model to the monitored dynamics of Majorana fermions later.

\begin{figure}[h]
    \centering
    \includegraphics[width=1\linewidth]{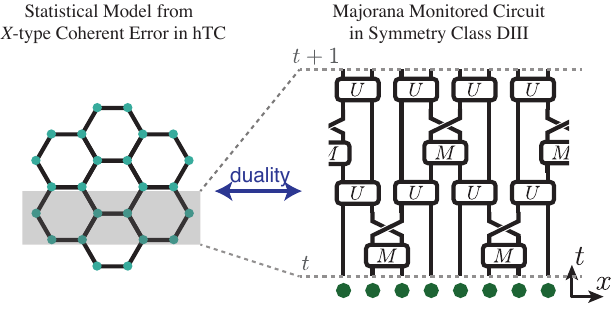}
    \caption{The disordered statistical model (left panel) that describes the syndrome distribution of the honeycomb-lattice toric code with $X$-type coherent error can be dualized to a 1+1D monitored circuit (right panel) acting as a chain of Majorana fermion modes (green dots). This Majorana monitored circuit is non-interacting and belongs to the AZ symmetry class DIII. The shaded part of the statistical model is mapped to the circuit block between time $t$ and $t+1$. Unitary gates and measurements are labeled by $U$ and $M$, respectively.}
    \label{fig:MajCircuit_hTCX}
\end{figure}

Next, we show that the partition function ${\cal Z}_{X,\{\eta_l\}}$ of the disordered statistical model is dual to the quantum amplitude of a 1+1D dynamical Majorana chain evolving under a monitored quantum circuit (see Fig. \ref{fig:MajCircuit_hTCX}). We summarize the dual Majorana monitored circuit below and leave the detailed derivation in App. \ref{app:mapcirc}. Let $\psi_n$ denote the Majorana fermion operator on the $n$th site of the Majorana chain. Every horizontal link $l$ in the statistical model is mapped to a generalized measurement in the monitored circuit. This measurement combines a weak measurement of the local fermion parity $\ii \psi_n\psi_{n+1}$ on two neighboring sites and a (fermionic) SWAP gate. In this context, $\eta_l = \pm 1$ corresponds to the two measurement outcomes. A set ${\cal M}_l$ of Kraus operators, one for each outcome, determines how wavefunction evolution (or collapse) under the measurement. The graphical representation and the Kraus operator set ${\cal M}_l = \{M_{\eta_l=\pm1}\}$ associated with this generalized measurement are given by 
\begin{align}
    \raisebox{-4mm}{\includegraphics[height=1cm]{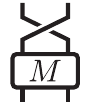}}
:~~ M_{\eta_l}=\frac{1}{\sqrt{2\cosh 2\kappa_l}}e^{- \eta_l \kappa_l \ii \psi_n \psi_{n+1} } e^{\eta_l \frac{\pi}{4}  \psi_n \psi_{n+1} }
\label{eq:KrausOp}
\end{align}
with the measurement strength $\kappa_l = \frac{1}{2}\log\tan\theta_l$. One can check that this Kraus-operator set ${\cal M}_l$ describes a physical measurement as it satisfies the condition for a positive operator-value measure\cite{nielsen2012quantum}: $\sum_{\eta_l = \pm 1} M_{\eta_l}^\dag M_{\eta_l} = \mathds{1}$. Physically, for a normalized incoming state $|\Psi\rangle$, the two measurement outcomes $\eta_l = \pm1$ also label the two associated quantum trajectories $|\Psi\rangle \rightarrow M_{\eta_l} |\Psi\rangle/||M_{\eta_l} |\Psi\rangle||$. The standard Born-rule probability for each quantum trajectory is given by $\langle\Psi|M_{\eta_l}^\dag M_{\eta_l}|\Psi\rangle$. We remark that it is a non-trivial fact that the different values of $\eta_l$ in the statistical models can be mapped to outcomes of a generalized measurement in the monitored circuit, rather than merely a random parameter in a generic non-unitary evolution.

Every non-horizontal ($60^\circ$ or $120^\circ$) link in the statistical model is mapped to a nearest-neighbor random unitary gate $U_{\eta_l}$ in the Majorana circuit:
\begin{align}
    \raisebox{-4mm}{\includegraphics[height=1cm]{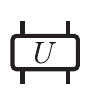}}
:~~ U_{\eta_l} = e^{-\left[\theta_l + \frac{\pi}{4}(1-\eta_l) \right]\psi_n \psi_{n+1}}
\label{eq:RandUnitary}
\end{align}
$\eta_l$ should be treated as a random variable that takes $\pm 1$ with equal probability. One can also interpret $\eta_l$ as a label of quantum trajectories under the random unitary evolution $U_{\eta_l}$. 

With every link in the statistical model mapped to either a measurement or a random unitary, the entire statistical model is dualized to a monitored quantum circuit on a Majorana chain as shown in Fig. \ref{fig:MajCircuit_hTCX}. All the links in the shaded areas of the statistical models (left panel) are mapped to the circuit block (right panel) between time steps $t$ and $t+1$ in the Majorana monitored dynamics. In the spatial direction, the monitored circuit has a four-site unit cell on the Majorana chain. In the temporal direction, the same circuit block repeats in each unit time step.

The variables $\{\eta_l\}$, which denote the random bond configuration in the statistical model, are now mapped to quantum trajectories in the Majorana monitored circuit. The partition function ${\cal Z}_{X,\{s_p\}}$ can be viewed as the quantum amplitude generated by the product of random unitaries and the Kraus operators associated with the measurements. The Born-rule probabilities of the quantum trajectories are exactly given by $|{\cal Z}_{X,\{\eta_l\}}|^2$. Hence, singularities in the disorder-average free energy $F_X$ [Eq. \eqref{eq:hTCX_FreeEnergy}], which mark the decodability transition in the hTC, also signal the ``measurement-induced" transitions between dynamical phases of the Majorana monitored circuit.

A standard analytical treatment of $F_X$ utilizes the replica trick: $F_X \propto \lim_{R\rightarrow 1} \frac{-1}{R-1}\log \left( \sum_{\{\eta_l\}} |{\cal Z}_{X,\{\eta_l\}}|^{2R} \right) + \text{constant}$. Physically, $|{\cal Z}_{X,\{\eta_l\}}|^{2R}$ can be viewed as the quantum amplitude of the $R$-replica verion of the monitored circuit and its conjugate, which are the circuits responsible for the evolution of $R$ replicas of the density matrix of the Majorana chains. It is important to note that the replica limit $R\rightarrow 1$ needs to be taken at the end to properly understand the decodability of hTC and the properties of the dual Majorana monitored circuit. We remark that a similar replica limit $R\rightarrow 1$ also arises in the case of toric codes with incoherent error due to the Nishimori condition in the associated disordered statistical model \cite{Dennis2002topo}. However, the content of each replica differs significantly from the current case.

\subsubsection{Symmetry Class, Possible Phases, and Continuum Description}
\label{sec:SymmetryClass_hTCX}

Notably, the Majorana monitored circuit dual to the decoding problem of the hTC with $X$-type coherent error is {\it non-interacting}: In any given quantum trajectory (labeled by $\{\eta_l\}$), the Kraus operators $M_{\eta_l}$ and the unitary gates $U_{\eta_l}$ [Eqs. \eqref{eq:KrausOp} and \eqref{eq:RandUnitary}] in the circuit are both exponentiated fermion bilinears.
There has been extensive prior research on the monitored dynamics of non-interacting fermions. An important result is that the universal properties of the non-interacting monitored circuit are dictated by its AZ symmetry 
class\cite{zirnbauer1996riemannian,altland1997nonstandard,SchnyderRyu2010NJP,JianRTN2022,jian2023measurement}. In the current case, the Majorana monitored circuit has no symmetry other than the global fermion parity and, consequently, belongs to symmetry class DIII. This is to be contrasted with the case of $Z$-type coherent error discussed later in Sec. \ref{sec:hTCZ} and the square-lattice case in Sec. \ref{sec:sTC}.

Class-DIII non-interacting Majorana monitored circuits in 1+1D have been extensively investigated both numerically and analytically in the literature (see examples in Refs. \onlinecite{JianRTN2022,FavaNahum2023,kells2023topological,jian2023measurement,Negari2024,GYZhuLearning2025}). We will first briefly review the existing results and then study their implications for hTC with $X$-type coherent error. The class-DIII non-interacting Majorana monitored circuits in 1+1D can exhibit two topologically distinct dynamical phases with area-law entanglement scaling and a critical phase with logarithmic entanglement scaling. The critical phase is sometimes also referred to as the ``metallic phase" because of its close connection to the disordered thermal metal phase of the class-DIII Anderson localization problem in 2 spatial dimensions\cite{JianRTN2022,Fulga2012DIIIMIT,Venn_2023}.

These three phases can be unified under a continuum field-theory description given by a 2D non-linear sigma model (NSLM)\cite{JianRTN2022,jian2023measurement,FavaNahum2023}. This NLSM captures the long-wavelength behavior of the $R$-replica monitored circuit. Its action is 
\begin{align}
    {\cal S}_{\rm DIII} = \int dt dx \frac{1}{g_3}{\rm Tr}\left( \partial_t O^T \partial_t O + \partial_x O^T \partial_x O \right), 
    \label{eq:NLSMhTC}
\end{align}
where $t$ and $x$ refer to the temporal and spatial coordinates of the circuit. $g_3$ is the coupling constant that is tied to the coarse-grained measurement strength. The dynamical degree of freedom $O$ is a $R\times R$ matrix field that takes value in the target manifold $\SO(R)$. This target manifold is fully dictated by the symmetry class DIII. The global symmetry of this NLSM is $\SO(R) \times \SO(R)$, which corresponds to the left and right multiplication of $O$ by $\SO(R)$ matrices $O_{+,-}$: $O \rightarrow O_+ O O_-^T$. Microscopically, the two $\SO(R)$ symmetries correspond to the continuous rotations amongst the $R$ replicas of the monitored circuits (whose quantum amplitude is ${\cal Z}_{X,\{\eta_l\}}$ in each replica) and those amongst the $R$ replicas of conjugate circuits (whose quantum amplitude is ${\cal Z}^*_{X,\{\eta_l\}}$ in each replica). 

The critical phase corresponds to the spontaneous continuous symmetry breaking of $\SO(R) \times \SO(R)$ to the diagonal $\SO(R)$ symmetry, which is stable even in 2 dimensions in the replica limit $R\rightarrow 1$\cite{jian2023measurement,FavaNahum2023,JianRTN2022}. At the ``saddle-point'' level, this phase is described by the vacuum expectation value $\langle O \rangle \propto I_{R\times R}$, where $I$ is the $R\times R$ identity matrix. Under the renormalization group (RG), the coupling constant flows to the weak coupling limit $g_3\rightarrow 0$. The Goldstone mode fluctuations in this phase give rise to logarithmic entanglement scaling and power-law fermion correlation functions in the long-time limit of the monitored dynamics. The $R\rightarrow 0$ counterpart of this critical phase exactly corresponds to the disordered thermal metal phase of the class-DIII Anderson localization problem in 2 spatial dimensions.

There are two phases with area-law entanglement scaling in the 1+1D Majorana monitored dynamics. They are characterized by the symmetric phases where the full symmetry $\SO(R) \times \SO(R)$ remains unbroken, and all the correlations are short-ranged. They occur at strong coupling $g_3\gtrsim1$. Phase transitions from the critical phases to these two area-law phases are driven by the proliferation of vortices, which are classified by $\pi_1(\SO(R)) = \mathbb{Z}_2$\cite{jian2023measurement,FavaNahum2023}. The two area-law phases are topologically distinct because of their different signs in the vortex fugacity. 

From the microscopic perspective of the monitored circuit, each area-law phase can be conceptually represented by a dimerization 
pattern on the Majorana chain.\cite{FavaNahum2023,nahum2020entanglement,kells2023topological,jian2023measurement}. The dimerization pattern is stabilized by repeated measurements in the circuit, which collapses the many-body wavefunction into pairs of entangled Majorana modes. The two topologically distinct dimerization patterns are reminiscent of the two gapped phases of the celebrated 1D Kitaev chain.\cite{Kitaev2001} The two area-law phases can be distinguished using a $\mathbb{Z}_2$-valued topological index ${\cal I}$, which counts the parity of the number of Majorana zero modes per entanglement cut in the entanglement spectrum of an (even-length) interval in the Majorana chain. Conceptually, ${\cal I}$ simply counts the parity of the number of Majorana dimers that cross a given cut in the Majorana chain. A different formulation of this topological index, based on the Lyapunov spectrum of the monitored dynamics, can be found in Refs. \onlinecite{OshimaTopo2025,BehrendsSurfaceCodesQuantumCircuits}.

Now, we return to the context of the hTC. All three phases of the 1+1D class-DIII Majorana monitored circuit should correspond to phases in the decodability phase diagram of the hTC with $X$-type coherent errors. One of the area-law dynamical phases in the Majorana monitored circuit corresponds to the decodable phase, while the other two phases correspond to undecodable phases of the hTC with different behaviors. 
As we will see in Sec. \ref{sec:two-parameter_hTCX}, all these phases and their phase transition can be reached within a two-parameter family of error angles $\{\theta_l\}$. 

Before diving into the details of the two-parameter model, one may ask whether it is too restrictive. In principle, each $\theta_l$ is an independent parameter and likely to vary in space in realistic quantum devices. However, it is important to note that the $\{\eta_l\}$ variables, which label the quantum trajectories, are already random variables in the monitored dynamics (and also the disordered statistical models). Additional randomness from the spatial variation in the error angle $\theta_l$ can affect the location of the decodability transition, but it will NOT change the universal properties of the phases in the decodability phase diagram or the nature of the phase transitions.

\begin{figure*}[t]
        \centering
    \includegraphics[width=1\linewidth]{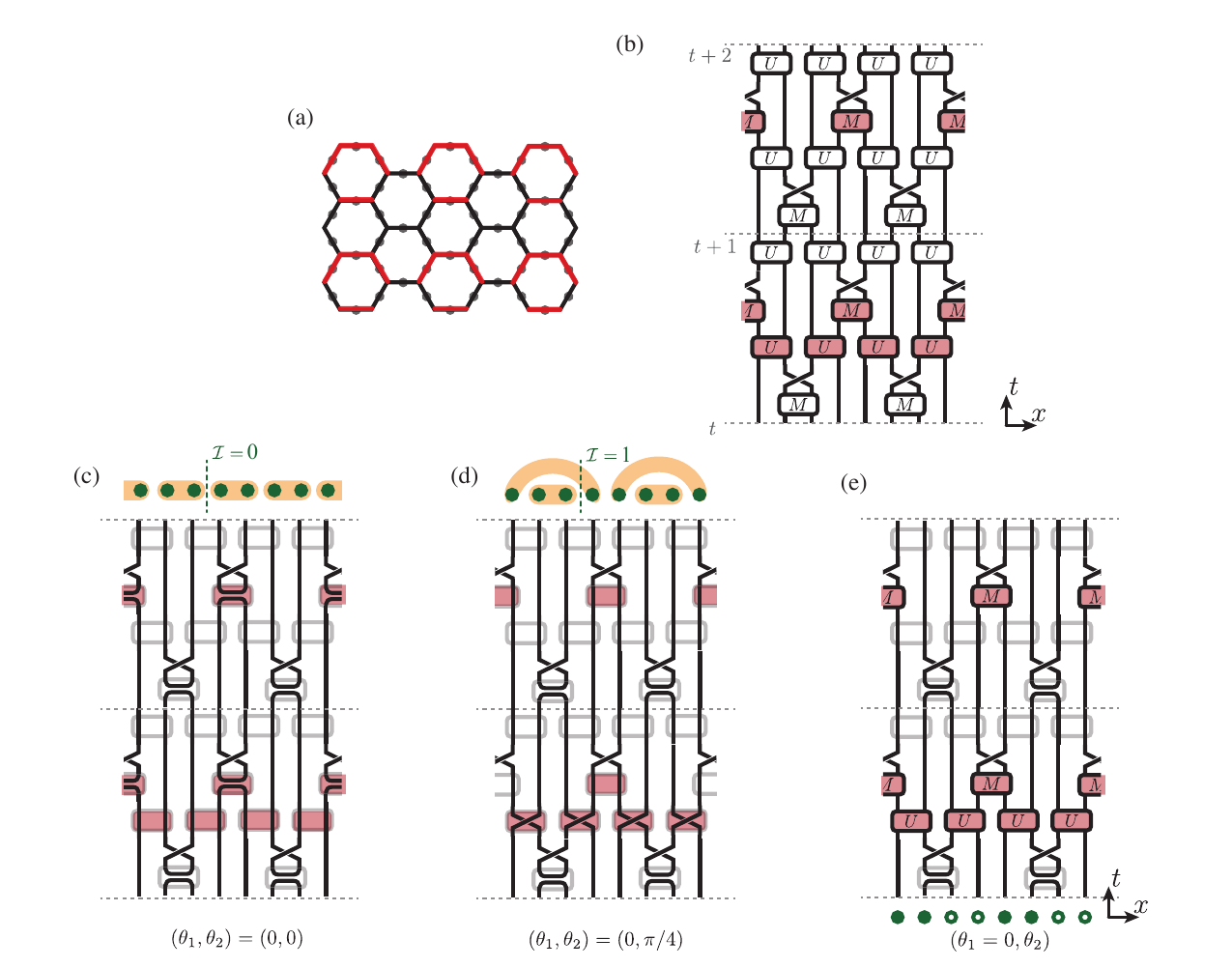}
    \caption{(a) The two-parameter coherent error model: $e^{\ii\theta_1 X_l}$ on the black links and $e^{\ii\theta_2 X_l}$ on the red links. (b) The Marjoana monitored circuit model associated with the two-parameter coherent error model. The red gates are associated with red links on the honeycomb lattice, and the white gates are associated with black links. The pattern of gates in the monitored circuit has a temporal periodicity of two unit time steps. (c) The Majorana monitored circuit with $(\theta_1,\theta_2)= (0,0)$ stabilizes the nearest-neighbor Majorana dimerization pattern shown at the top. This pattern has the topological index $\cI = 0$ with respect to the green dashed cut in the Majorana chain.
    (d) The Majorana monitored circuit with $(\theta_1,\theta_2)= (0,\pi/4)$ stabilizes a Majorana dimerization pattern (shown at the top) with topological index $\cI=1$. 
    (e) The Majorana monitored circuit with $(\theta_1=0,\theta_2)$ has an extra emergent TR symmetry. This TR symmetry is defined with respect to a sublattice structure in the Majorana chain. The solid green dots represent the $A$ sublattice, while the hollow green dots represent the $B$ sublattice. 
    }
    \label{fig:TwoParameter_hTCX}
\end{figure*}

\subsubsection{Two-parameter Coherent Error Model, Solvable Limits, and Emergent Symmetry}
\label{sec:two-parameter_hTCX}
We introduce a two-parameter model as a minimal model to explore the decodability phase diagram of the hTC with $X$-type coherent error. We divide the qubits of the hTC into two groups as indicated by the red and black links in Fig. \ref{fig:TwoParameter_hTCX}(a). The error angles are $\theta_l = \theta_1$ on the black links and $\theta_l = \theta_2$ on the red links. Even though this error pattern may seem ad hoc, it is a minimal model that captures all possible phases in the decodability phase diagram. As we argue at the end of Sec. \ref{sec:SymmetryClass_hTCX}, allowing further spatial variation beyond this two-parameter model can deform the location of the phase boundaries in the decodability phase diagram but will not alter the universal behavior of the phases and their transitions. 

We analyze the decodability phase diagram associated with the two-parameter error model using the dual Majorana monitored circuit. The black (red) links in the honeycomb lattice shown in Fig. \ref{fig:TwoParameter_hTCX}(a) map to the white (red) gates in the dual monitored circuit depicted in Fig. \ref{fig:TwoParameter_hTCX}(b). The white gates are associated with parameter $\theta_1$ while the red ones are associated with $\theta_2$. The pattern of the gates repeats every 2 unit time steps. The full decodability phase diagram can be obtained by numerically simulating the Majorana monitored circuit, which we will summarize in Sec. \ref{sec:hTCX_PhaseDiagram}. Before we discuss the full phase diagram, it is helpful to first study various analytically solvable points of the circuit and some general analytic features of this two-parameter model. 

We begin by discussing various limits of the generalized measurement ${\cal M}$ and the random unitary gate $U$ [defined in Eqs. \eqref{eq:KrausOp} and \eqref{eq:RandUnitary}]. We suppress the subscript $l$, i.e., the link label, for simplicity. For the generalized measurement $\cal M$, there are two limits 
\begin{align}
    \raisebox{-4mm}{\includegraphics[height=1cm]{Mgate.pdf}}
: M_{\eta} \rightarrow \begin{cases}
    \frac{1- \ii \eta \psi_n \psi_{n+1}}{2} e^{\eta \frac{\pi}{4} \psi_n \psi_{n+1}} & \text{for~} \theta = 0,\\
    e^{\eta \frac{\pi}{4} \psi_n \psi_{n+1}} &\text{for~} \theta = \pi/4.
\end{cases}
\end{align}
The two limits are graphically represented as $\raisebox{-3mm}{\includegraphics[height=0.8cm]{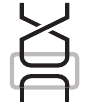}}$ for $\theta = 0$ and $\raisebox{-3mm}{\includegraphics[height=0.8cm]{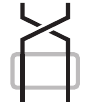}}$ for $\theta = \frac{\pi}{4}$. With $\theta = 0$, the measurement becomes projective. It forces the measured pair of Majorana modes to form a maximally entangled dimer (that decouples from all other Majorana modes). With $\theta = \pi/4$, the measurement reduces to a unitary SWAP gate between the neighboring Majorana modes. This graphical representation highlights the evolution of the entanglement patterns between the Majorana modes.  

For the unitary gate $U$, the two limits yield
\begin{align}
    \raisebox{-4mm}{\includegraphics[height=1cm]{Ugate.pdf}}
: U_{\eta} 
\rightarrow \begin{cases}
    (-\psi_n \psi_{n+1})^{\frac{1-\eta}{2}} & \text{for~} \theta = 0,\\
    (-\psi_n \psi_{n+1})^{\frac{1-\eta}{2}} e^{ -\frac{\pi}{4}  \psi_n \psi_{n+1}} &\text{for~} \theta = \pi/4.
\end{cases}
\end{align}
The two limits are graphically represented as $\raisebox{-3mm}{\includegraphics[height=0.8cm]{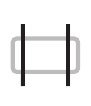}}$ for $\theta \rightarrow 0$ and $\raisebox{-3mm}{\includegraphics[height=0.8cm]{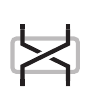}}$ for $\theta \rightarrow \frac{\pi}{4}$. This graphical representation mostly emphasizes the direction in which the Majorana modes propagate under $U_\eta$. The Majorana modes can gain additional phases through the factor $(-\psi_n \psi_{n+1})^{\frac{1-\eta}{2}}$.

The first solvable point is given by $(\theta_1, \theta_2) =(0,0)$, which is the limit without any coherent error. Obviously, this point represents the decodable phase. From the perspective of the Majorana monitored circuit, the measurements drive the Majorana chain into an area-law phase featuring a pattern of maximally entangled Majorana dimers shown on the top of Fig. \ref{fig:TwoParameter_hTCX}(c). This phase has the topological index ${\cal I}=0$ defined with respect to the cut shown as the green dashed line.

Another solvable points are given by $(\theta_1,\theta_2)= (0,\pi/4)$. One can easily see that the Majorana monitored circuit at this point represents an area-law phase featuring the Majorana dimerization pattern with ${\cal I} = 1$ shown in Fig. \ref{fig:TwoParameter_hTCX}(d). This area-law phase is topologically distinct from the decodable area-law phase.

Moreover, $(\theta_1, \theta_2) = (\pi/4, \pi/4)$ corresponds to the case where all the gates are essentially reduced to unitary SWAP gates on neighboring sites (modulo extra unitary operators of the form $(-\psi_n \psi_{n+1})^{\frac{1-\eta}{2}}$). At this special unitary point, the Majorana modes propagate ballistically. A small departure from this point will lead to a small but non-vanishing measurement strength in the generalized measurements. The entanglement entropy scaling in the neighborhood of $(\theta_1, \theta_2) = (\pi/4, \pi/4)$ is expected to follow the logarithmic entanglement scaling, which we confirm using numerical simulation (see Sec. \ref{sec:hTCX_PhaseDiagram}). 

In addition to these solvable points, our two-parameter error model has several convenient properties. First, one can show the error-corrupted states $\cEx$ with $(\theta_1 + n_1 \pi/2 , \theta_2 + n_2 \pi/2)$ are identical for all $n_1, n_2 \in \mathbb{Z}$. That is because the unitary rotations $\prod_l e^{\ii \theta_l X_l}$ with these parameters $(\theta_1 + n_1 \pi/2 , \theta_2 + n_2 \pi/2)$ differ from each other by products of the vertex stabilizers $A_v$ which act trivially on the logical state $|\Omega\rangle$.

Also, notice that under the TR symmetry $\cal T$ action Eq. \eqref{eq:TR_Error}, the parameters $(\theta_1,\theta_2)$ are mapped to $(-\theta_1,-\theta_2)$. Therefore, $(\theta_1,\theta_2)$ and $(-\theta_1,-\theta_2)$ should belong to the same phase in the decodability phase diagram (or equivalently, in the phase diagram of the Majorana monitored dynamics). Now, we can define a ``fundamental domain" with $\theta_1 \in [-\pi/4,\pi/4]$ and $\theta_2 \in [0,\pi/4]$ of the decodablity phases diagram parameterized by $(\theta_1,\theta_2)$. Any $(\theta_1,\theta_2)$ can be related to a point in the fundamental domain via the periodicity of $\theta_{1,2}$ and the TR symmetry action $\cT$.

Another interesting feature of our two-parameter error model is an emergent TR symmetry $\cT_e$ in the Majorana monitored circuit with $\theta_1 = 0$ [see Fig. \ref{fig:TwoParameter_hTCX}(e)]. This emergent TR symmetry comes from an additional sublattice structure of the Majorana chain. The two sublattices $A$ and $B$ are depicted as the solid and hollow green dots in Fig. \ref{fig:TwoParameter_hTCX}(d). This emergent TR symmetry $\cT_e$ is an anti-unitary action that transforms the Majorana modes as 
\begin{align}
    \cT_e: \begin{cases}
        \psi_n \rightarrow \psi_n,~~& n\in \text{sublattice~} A, \\
        \psi_n \rightarrow -\psi_n,~~& n\in \text{sublattice~} B,
    \end{cases} \label{eq:TRe}
\end{align}
Notice that each unitary gate $U_{\eta_l}$ in the monitored circuit acts only within a sublattice. One can show that such unitary gates always commute with $\cT_e$:
\begin{align}
    \cT_e U_{\eta_l} \cT_e^{-1} = U_{\eta_l}.
\end{align}
On the other hand, the generalized measurements ${\cal M}_l$ always act across different sublattices. The Kraus operator transformations under $\cT_e$ as 
\begin{align}
    \cT_e M_{\eta_l} \cT_e^{-1} =  (\ii)^\eta (\ii \psi_n \psi_{n+1}) M_{\eta_l}
\end{align}
The extra operator $(\ii)^\eta (\ii \psi_n \psi_{n+1})$ on the righthand side breaks the $\cT_e$ symmetry for generic parameters $(\theta_1,\theta_2)$. However, when $\theta_1=0$, every measurement ${\cal M}_l$ is either immediately followed by or preceded by a measurement at the same location. Two consecutive measurements ${\cal M}_{l_{1,2}}$ on the same pair of Majorana modes commute with the $\cT_e$ operation (upto an unimportant global phase):
\begin{align}
    \cT_e M_{\eta_{l_1}} M_{\eta_{l_2}} \cT_e^{-1} \propto M_{\eta_{l_1}} M_{\eta_{l_2}}.
\end{align}
The extra local fermion parity operators $(\ii \psi_n \psi_{n+1})$ from each measurement cancel out. 

When the Majorana monitored circuit commutes with an extra TR symmetry $\cT_e$ obeying ${\cal T}_e^{2}=1$, its AZ symmetry class becomes D (instead of DIII)\cite{bhuiyan2025free}. In 1+1D, symmetry-class-D non-interacting Majorana monitored circuit can exhibit a $\mathbb{Z}$-classified set of topologically distinct area-law phases\cite{bhuiyan2025free,Pan_2025,XiaoKawabata2024,SchnyderRyu2010NJP}. The ``even class" and the ``odd class" reduce to the two types of class-DIII area-law phases once the extra TR symmetry is explicitly broken. More interestingly, the critical phase (or the ``metallic phase") is expected to be absent in 1+1D class-D Majorana monitored dynamics. Later, we will see that symmetry class D also appears in hTC with $Z$-type coherent errors and the square-lattice toric code with coherent errors, but without any fine-tuning of parameters. We refer to Sec. \ref{sec:SymmetryClass_hTCZ} for a more detailed discussion on the possible area-law phases and the absence of a critical phase in 1+1D class-D monitored dynamics.

\subsubsection{Phase Diagram}
\label{sec:hTCX_PhaseDiagram}

\begin{figure*}[t!]
    \centering
    \includegraphics[width=0.85\linewidth]{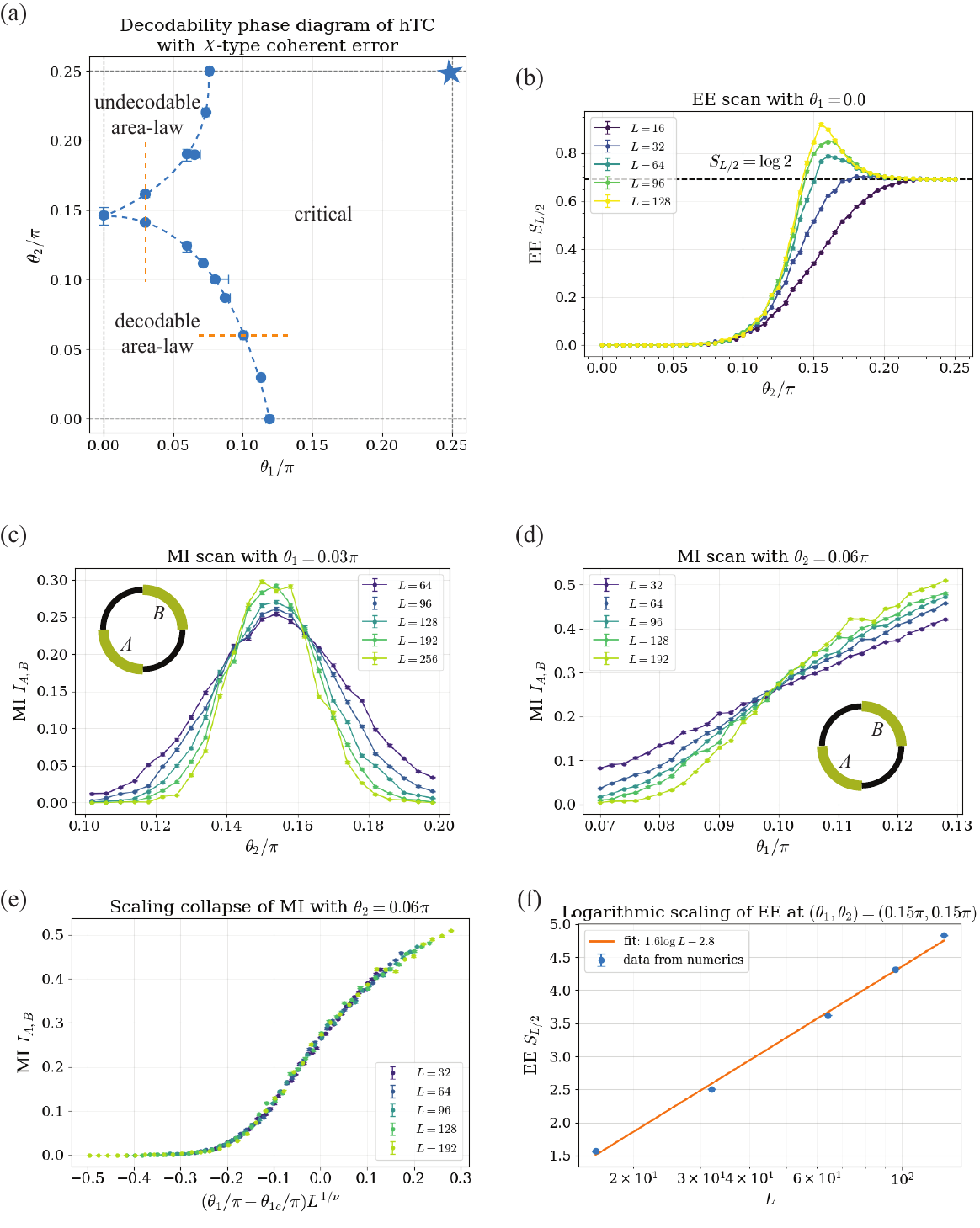}
    \caption{(a) Decodability phase diagram obtained from the numerical simulations of the dual Majorana monitored circuit. 
    (b) Half-system entanglement entropy (EE) $S_{L/2}$ along the line $\theta_1=0$. 
    Two area-law phases are separated by a phase transition at $\theta_2\approx 0.146\pi$.    
    (c)(d) Representative scans along 1D slices on the phase diagram (dashed orange lines in (a)) of the mutual information (MI) of two antipodal quarter-system-size intervals $A$ and $B$ (see inset of (c)). The phase boundaries are determined by MI crossings along multiple slices of the phase diagram 
    (e) Scaling collapse of the MI data $I_{A,B}$ along the line $\theta_2 = 0.06\pi$ shown in (d).
    (f) Logarithmic EE scaling at a typical point in the critical phase. 
    The MI data are averaged over 5000 samples, while the EE data are averaged over 2000 samples.  }
    \label{fig:honeyphase}
\end{figure*}

The phase diagram for the two-parameter coherent error model for the hTC is plotted in Fig. \ref{fig:honeyphase}(a). Because of the aforementioned relations 
\begin{equation*}
(\theta_1,\theta_2)\sim (\theta_1+n_1\pi/2, \theta_2+n_2\pi/2)\sim(-\theta_1, -\theta_2),
\end{equation*}
between different $\twotheta$'s of the monitor circuit (see Sec. \ref{sec:SymmetryClass_hTCX}), a fundamental domain of the phase diagram is
given by the range $-\pi/4\leq \theta_1\leq \pi/4$ and $ 0 \leq \theta_2 \leq \pi/4$. We observe a similarity between the phase diagram in the positive and negative $\theta_1$ regimes. Hence, we focus on the phase diagram within the regime $0\leq\theta_{1,2}\leq\pi/4$.

The phase diagram is obtained by numerically simulating the monitored circuit shown in Fig. \ref{fig:TwoParameter_hTCX}(b) on periodic Marjoana chains of various lengths $L$. The spacetime aspect ratio of the circuit is fixed. Since this monitored circuit is non-interacting, the numerical simulation can be carried out efficiently using fermionic Gaussian circuit methods (See App. \ref{app:algocircuit} for more details on the numerical method). We use the entanglement properties of the final states of the circuit (averaged over many quantum trajectories) to distinguish different phases and locate phase transitions. 

The phase diagram we obtain is shown in Fig. \ref{fig:honeyphase}(a). It contains a decodable area-law phase, an undecodable area-law phase, and an undecodable critical phase. These are all possible phases in symmetry class DIII as discussed in Sec. \ref{sec:SymmetryClass_hTCX}. Here, by ``(un)decodable", we refer to the corresponding phase in the error-corrupted hTC. We calculate the average half-system entanglement entropy (EE) $S_{L/2}$ on the Majorana chain. We confirm that EE $S_{L/2}$ saturates to an ${\cal O}(1)$ value as the $L$ increases in each area-law phase. An example is shown in Fig. \ref{fig:honeyphase}(b) for $\theta_1 = 0$ where the two area-law phases are separated by a direct phase transition at $\theta_2 \approx 0.146\pi$. These two area-law phases are represented by the two dimerization patterns [shown in Fig. \ref{fig:TwoParameter_hTCX} (c-d)] with $\cI=0$ and $\cI=1$ in the limits $(\theta_1,\theta_2) = (0,0), (0,\pi/4)$. Fig. \ref{fig:honeyphase}(f) provides an example $(\theta_1, \theta_2) = (0.15\pi,0.15\pi)$ for the critical phase where the EE scaling follows $S_{L/2} \sim \log L$. Technically, the expected entanglement scaling in the critical phase contains an extra $(\log L)^2$ correction\cite{FavaNahum2023}. However, reliably extracting this correction from the numerical simulation likely requires a much larger range of system sizes. As discussed in Sec. \ref{sec:SymmetryClass_hTCX}, $(\theta_1,\theta_2)=(\pi/4,\pi/4)$ is a special point (indicated by the star in Fig. \ref{fig:honeyphase}) where the monitored circuit reduces to a collection of unitary SWAP gates. With a random, pure initial state for the circuit, we expect volume-law EE scaling in the final state, which we confirm numerically.

From the hTC perspective, the decodable area-law phases grow out of the error-free point $\twotheta=(0,0)$. Transition out of this phase should indicate the loss of decodability. The limit $\twotheta=(0,\pi/4)$ represents the undecodable area-law phase represented by the $\cI=1$ dimerization pattern. When a qubit $l$ of a hTC logical state is rotated by $e^{\ii \frac{\pi}{4} X_l} = \frac{1+\ii X_l}{\sqrt{2
}}$, the wavefunction becomes an equal weight position between the state with a pair of non-trivial syndromes on the plaquettes that border $l$ and the state without. When all the red links have an error angle $\theta_2 =\pi/4$, the syndromes can coherently propagate through the entire lattice (even when $\theta_1 =0$), rendering the error-corrupted state undecodable. This physical picture agrees with our conclusion that the area-law phase around $\twotheta=(0,\pi/4)$ is undecodable. By similar reasoning, the critical phase surrounding $\twotheta=(\pi/4,\pi/4)$ is also undecodable. In this work, we focus on locating these three phases and their transitions in the decodability phase diagram of hTC and understanding how the symmetry class affects their relative positions. Systematically studying how the logical error rate varies with the hTC's code distance requires more careful treatment of the code space, which depends on the boundary conditions or the spatial topology of the hTC. We will defer this investigation for future work.

To pin down the phase boundaries and extract the critical behavior, we calculate the mutual information (MI) $I_{A,B}$ in the final state of the monitored circuit between two antipodal intervals $A$ and $B$ with length $L/4$. $L$ is the length of the Majorana chain. For example, a pair of such antipodal intervals is given by $A=[x_0, x_0+L/4]$ and $B= [x_0+L/2, x_0+3L/4]$. At the phase transitions, this antipodal MI $I_{A,B}$ is expected to be independent of system size due to the emergence of conformal symmetry \cite{LiConformal2021}. Fig. \ref{fig:honeyphase} (c-d) shows how this antipodal MI $I_{A,B}$ depends on $\twotheta$ along to two example slices of the phase diagram. The MI crossings for different system sizes mark the location of the phase transition. In Fig. \ref{fig:honeyphase} (e), we present a scaling collapse of the MI data shown in Fig. \ref{fig:honeyphase} (d). Based on the scaling collapse, we find the critical value $\theta_{1c}/\pi = 0.1000^{+0.0016}_{-0.0019}$ for the decodability transition along the line with $\theta_2 =0.06\pi$. The fitted correlation-length exponent is $\nu=2.28^{+0.17}_{-0.23}$, which is consistent with the previously studied area-law-to-critical phase transition of class-DIII non-interacting Majorana monitored circuits\cite{jian2023measurement}. The algorithm for the scaling collapse is explained in App. \ref{app:algo_ScalingCollapse}

The decodability phase diagram in Fig. \ref{fig:honeyphase}(a) contains important features that highlight the role of symmetry classes in controlling the structure of the phase diagram. In class-DIII non-interacting Majorana monitored circuits, the two area-law phases are generically separated by the critical phase. This feature has been seen in many previous studies (for example, Refs.~\onlinecite{kells2023topological,Pan_2025,FavaNahum2023, Negari_2024}) and is consistent with our phase diagram Fig. \ref{fig:honeyphase}(a) for $\theta_1 >0 $. As we've shown in Sec. \ref{sec:SymmetryClass_hTCX}, when $\theta_1 =0$, the Majorana monitored circuit has an extra emergent TR symmetry $\cT_e$ that shifts the system's symmetry class to D (from DIII). In symmetry class D, the critical phase is no longer stable under RG (see Sec. \ref{sec:SymmetryClass_hTCZ}). Hence, the two area-law phases can be adjacent to each other, separated by a direct continuous phase transition, which is also manifested in the phase diagram Fig. \ref{fig:honeyphase}(a). These results show that the nature of the transitions out of the decodable phase is indeed controlled by the AZ symmetry class.

\subsection{$Z$-type coherent error}
\label{sec:hTCZ}

In the following, we study the hTC subject to $Z$-type coherent errors. Similar to the previous discussion, we introduce a two-dimensional disordered statistical model and its dual description in terms of 1+1D Majorana monitored dynamics, which together capture the syndrome distribution $P_Z({s_v})$. We show that, in this case, the dual Majorana dynamics belong to symmetry class D (in contrast to class DIII for the $X$-type error), and we elucidate how this symmetry class emerges from a microscopic time-reversal symmetry. We further discuss the possible phases and RG behavior within this symmetry class. We note that the square-lattice toric code subject to coherent errors also maps to class-D Majorana monitored dynamics; we will therefore use it, rather than hTC, as a representative example to numerically investigate a concrete decodability phase diagram in class D.

\subsubsection{Statistical Model and Majorana Monitored Circuit}

For hTC subject to $Z$-type coherent errors, the syndrome distribution ${\cal P}_Z(\{s_v\})$ in $|{\cal E}_Z\rangle$ can be written in terms of the partition function ${\cal Z}_Z(\{s_v\})$ of a classical disordered statistical model on a triangular lattice,  which is the dual of the orginal honeycomb lattice of the hTC (as shown in Fig. \ref{fig:honeycombTC}(c)):
\begin{equation}
     {\cal P}_Z(\{s_v\}) = |{\cal Z}_{Z,\{s_v\}}|^2.
\end{equation}
In the statistical model, classical Ising spins $\tau_p =\pm 1$ live on the vertices of the triangular lattice, which are dual to the plaquettes on the original honeycomb lattice. Hence, we use the plaquette label $p$ to denote the locations of $\tau_p$. The partition function ${\cal Z}_{Z,\{s_v\}}$ is given by
\begin{align}
    {\cal Z}_{Z,\{s_v\}} = {\cal N} \sum_{\{\tau_p =\pm1\}} \exp{\left[ - \sum_{\langle p,p'\rangle} J_{p,p'} \eta_{p,p'} \tau_p \tau_{p'} \right]}.
    \label{eq:hTCZ_DisorderedPartitionFunction}
\end{align}
$J_{p,p'}$ is a complex nearest-neighbor Ising coupling satisfying $e^{2J_{p,p'}} = \ii 
\tan\theta_l$ for the pair of neighboring plaquettes $p$ and $p'$ sharing the link $l=\langle p,p'\rangle$. The variable $\eta_{p,p'} = \pm 1$ captures the random-bond disorder on the triangular lattice. They are related to the syndromes via $\prod_{v\in l} \eta_l = s_v$. Different $\{\eta_l\}$ configurations that correspond to the same syndrome are gauge equivalent and yield the same partition function. ${\cal Z}_{Z,\{s_v\}}$ will also be written as ${\cal Z}_{Z,\{\eta_l\}} $ to highlight its dependence on the random bond configuration $\{\eta_l\}$. Just like the previous case, the disorder-averaged free energy $F_Z$ of this statistical model is given by 
\begin{align}
   F_Z \propto  &\sum_{\{\eta_l\}} -|{\cal Z}_{Z,\{\eta_l\}}|^2\log |{\cal Z}_{Z,\{\eta_l\}}|^2.
\end{align}
Singularities in $F_Z$ signal the transitions in the codes' intrinsic decodability.

\begin{figure}
    \centering
    \includegraphics[width=0.95\linewidth]{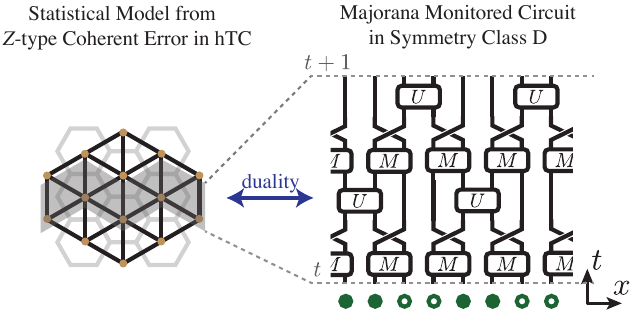}
    \caption{The disordered statistical model (left panel) that describes the syndrome distribution of the honeycomb-lattice toric code subject to $Z$-type coherent errors can be dualized to a 1+1D monitored circuit (right panel) acting on a Majorana chain. The Majorana chain has two sublattices $A$ (solid green dots) and $B$ (hollow green dots). This Majorana monitored circuit is non-interacting and belongs to AZ symmetry class D. The shaded part of the statistical model is mapped to the circuit block between times $t$ and $t+1$. Unitary gates and measurements are labeled by $U$ and $M$, respectively.}
    \label{fig:MajCircuit_hTCZ}
\end{figure}

Similar to the case of $X$-type error, this triangular-lattice disordered statistical model can be dualized to a Majorana monitored circuit as depicted in Fig. \ref{fig:MajCircuit_hTCZ}. The shaded area in the statistical model is mapped to the circuit block between the time $t$ and $t+1$. In particular, every vertical link on the triangle lattice is mapped to a random unitary gate $U_{\eta_l}$ as defined in Eq. \eqref{eq:RandUnitary}. Every non-vertical ($30^\circ$ and $150^\circ$) link on the triangle lattice is mapped to a generalized measurement ${\cal M}_l$ as defined in Eq. \eqref{eq:KrausOp}. Note that every link on the triangular lattice is dual to a link $l$ on the original honeycomb lattice. Hence, the $l$ naturally labels the gates in the monitored circuit.

This Majorana monitored circuit associated with the $Z$-type coherent error is again {\it non-interacting}. It has a four-site unit cell along the spatial direction. The circuit block shown in the right panel of Fig. \ref{fig:MajCircuit_hTCZ} repeats between every unit time step. 

\subsubsection{Symmetry Class, Possible Phases and Continuum Description}
\label{sec:SymmetryClass_hTCZ}

Recall that the error-corrupted state $|{\cal E}_Z \rangle$ respects the TR symmetry $\cal T$, as shown in Eq. \eqref{eq:TR_Error}. It is natural to expect an anti-unitary TR symmetry ${\cal T}_c$ in the resulting Majorana monitored circuit. To introduce this TR symmetry $\cT_c$, we divide the Majorana chain into two sublattices, $A$ and $B$, which are depicted as the solid and hollow green dots, respectively, in the right panel of Fig. \ref{fig:MajCircuit_hTCZ}. $\cT_c$ acts on the Majorana modes as
\begin{align}
    \cT_c: \begin{cases}
        \psi_n \rightarrow \psi_n,~~& n\in \text{sublattice}~A, \\
        \psi_n \rightarrow -\psi_n,~~& n\in \text{sublattice}~B.
    \end{cases}
\end{align}
Note all the unitary gates $U_{\eta_l}$ act within a sublattice, while all the measurements ${\cal M}_{\eta_l}$ act across different sublattices. Consequently, we have
\begin{align}
    & {\cal T}_c U_{\eta_l} {\cal T}_c^{-1} = U_{\eta_l}, \nonumber \\
    & {\cal T}_c M_{\eta_l} {\cal T}_c^{-1} =  (\ii)^\eta (\ii \psi_n \psi_{n+1}) M_{\eta_l}.
    \label{eq:TRonUM_hTCZ}
\end{align}
As shown in Fig. \ref{fig:MajCircuit_hTCZ}, the circuit block between time $t$ and $t+1$ contains two rows of measurements. Based on Eq. \eqref{eq:TRonUM_hTCZ}, when this circuit block is conjugated by $\cT_c$, each row of measurements produces an extra global fermion parity operator $\prod_n (\ii \psi_{2n-1}\psi_{2n})$ (modulo unimportant global phase factors), while the unitary gates stay intact. The two extra global fermion parity operators $\prod_n (\ii \psi_{2n-1}\psi_{2n})$ within each circuit block cancel each other. Therefore, as a whole, the circuit block in Fig. \ref{fig:MajCircuit_hTCZ} is invariant under the conjugation of $\cT_c$. In other words, the Majorana monitored circuit associated with the $Z$-type coherent error is TR symmetric. Since $\cT_c^2= 1$, this TR-symmetric Majorana monitored circuit belongs to symmetry class D\cite{bhuiyan2025free} (instead of DIII). We remark that our approach, which determines the symmetry class from the explicit symmetries of the error-corrupted state and the monitored circuit, is readily generalizable to broader classes of coherent errors, including those that render the Majorana monitored circuit interacting.

Notice the formal similarity between the TR symmetry $\cT_c$ for $Z$-type coherent errors and the emergent TR symmetry $\cT_e$ for a special class of $X$-type coherent errors discussed in Sec. \ref{sec:two-parameter_hTCX}. We emphasize that the former does not require any fine-tuning of the error angles $\theta_l$, while the latter does. This is expected as the $\cT_c$ is a manifestation of the TR symmetry of the $Z$-type error-corrupted state $|{\cal E}_Z\rangle$ for any values of $\{\theta_l\}$.  

Having established that the 1+1D Majorana monitored circuit arising from decoding the hTC with $Z$-type coherent errors belongs to symmetry class D, we now turn to the possible phases of this monitored circuit, which are dual to the phases in the corresponding decodability phase diagram. First, we discuss the area-law phases in the monitored circuit. In symmetry class D, the non-interacting area-law dynamical phases in 1+1D are classified by an integer 
$\mathbb{Z}$~ \cite{Pan_2025,XiaoKawabata2024,bhuiyan2025free,ClassificTopInsPRB2008,KitaevPeriodTable_2009,SchnyderRyu2010NJP}.
Each area-law phase can be represented by a dimerization pattern of the Majorana modes, which is stabilized by the repeated measurements in the circuit. The $\mathbb{Z}$ classification can be conceptually understood as follows. In the long-time limit, only the inter-sublattice dimers can be stabilized as they are invariant under $\cT_c$. Imagine we cut the (infinite) Majorana chain into two halves, break the Majorana dimers across the cut, and focus on the right half of the chain. A $\mathbb{Z}$-valued topological index ${\cal I}$ is given by the number of dangling Majorana modes on the $A$ sublattice minus the number of dangling Majorana modes on the $B$ sublattice. 

These area-law phases can also be captured by a NLSM,
which is the continuum description of the universal behaviors of the $R$-replica class-D monitored circuit in 1+1D. The action of this NLSM is given by
\begin{align}
    {\cal S}_{\rm D} = \int dt dx \frac{1}{g}{\rm Tr}\left( \partial_t Q^T \partial_t Q + \partial_x Q^T \partial_x Q \right) + \Theta \text{-term}. \label{eq:NLSMsTC}
\end{align}
$g$ is the coupling constant, which conceptually corresponds to the coarse-grained measurement strength. The matrix field $Q$ lives on the target manifold $\SO(2R)/\U(R)$, which is dictated by the symmetry 
class D~\cite{zirnbauer1996riemannian,SchnyderRyu2010NJP,jian2023measurement,FavaNahum2023}
\footnote{
The decodability transition in the presence of {\it incoherent} errors is known~\cite{Dennis2002topo} to be described by the 2D Random Bond Ising model at its critical point along the Nishimori line. A field theory for this transition was developed in Ref. \onlinecite{IlyaReadLudwig2001} as a special type of 2D Anderson localization problem of non-interacting Majorana fermions in class D, which is described by a universality class different from the one occurring in the circuit described in the present work. Our numerical results for the critical exponent $\nu$ at the transition in our circuit 
(Sec.~\ref{sec:hTCX_PhaseDiagram}, in particular Fig.~\ref{fig:squarephase}(d)) reflect this and are those~\cite{wang2025selfdual} of the NLSM in Eq.~(\ref{eq:NLSMsTC}).}.
The NLSM admits a $\Theta$-term due to $\pi_2(\SO(2R)/\U(R)) = \mathbb{Z}$. The symmetry of this NSLM is $\SO(2R)$. The behavior of the monitored circuit, as well as the decodability of the $Z$-type coherent error in hTC, is controlled by the replica limit $R\rightarrow 1$ of this NLSM. Alternatively, the replica limit $R\rightarrow 0$ of this NLSM describes the class-D Anderson localization problem in 2 spatial
dimensions~\cite{ChalkerKagalovskyEtAlThermalMetalPRB2001,SenthilFisher2000,IlyaReadLudwig2001,ReadLudwig2000,BocquetSerbanZirnbauerDirtySC2000,Fendley2001}. The $\mathbb{Z}$-classified area-law phases of the class-D Majorana monitored dynamics correspond to the strong-coupling RG fixed points of the NLSM, where $g$ flows to strong coupling, $Q$ has a vanishing vacuum expectation value, and the $\Theta$-angle flows to
$\Theta \in 2\pi \mathbb{Z}$.

The weak coupling limit with $g\rightarrow 0$ would be a critical or metallic phase with spontaneous continuous symmetry breaking. However, this critical phase (or metallic phase) turns out to be unstable under RG in the limit of 
$R\rightarrow1$~\cite{IlyaReadLudwig2001,FavaNahum2023,wang2025selfdual}.
For a fixed replica number $R$, the RG flow of this symmetry-class-D NLSM near the weak coupling limit is well known~\cite{Hikami1981,IlyaReadLudwig2001}
to read (up to 2-loop order)
\begin{align}
    \frac{dg}{d \log l } = 2(R-1) g^2 + 2 (R^2-3R+4) g^3+ {\cal O}(g^4).
    \label{eq:RG_ClassD}
\end{align}
Notice that when $R\rightarrow 1$, $\frac{dg}{d \log l } = 12 g^3 + \dots$,
causing $g$ to flow to strong coupling at long distances. Therefore, there is no weak-coupling critical phase in the class-D monitored circuit with $R\rightarrow 1$.~\cite{IlyaReadLudwig2001}

In the absence of a weak-coupling critical phase in the limit $R\rightarrow1$, it is possible that the $\mathbb{Z}$-classified area-law phases are the only stable phases of class-D monitored dynamics in 1+1D. We will treat this as our assumption even though we cannot completely rule out the possibility of non-area-law phases in the strong-coupling regime.

This RG equation with $R\rightarrow 1$ also suggests that the correlation length in the monitored circuit follows the asymptotic behavior $\xi \sim a \, e^{6/g^2_0}$ for a small bare coupling (or coarse-grained measurement strength) $g=g_0$. Here, $a$ is the lattice spacing. This asymptotic behavior indicates that the correlation length can be very large if the bare coupling $g_0$ is small. In a finite system whose size is much smaller than $\xi$, the monitored circuit, as well as the corresponding decoding properties, can behave like a critical phase even though the true long-distance behaviors are expected to follow those of an area-law phase. 
This phenomenon is potentially relevant in the neighborhood of the point where $\theta_l = \pi/4$ on all links $l$. When $\theta_l = \pi/4$ for all $l$, all the gates in the monitored circuit essentially reduce to unitary SWAP gates and, consequently, all Majorana modes evolve ballistically. A small deviation from this point can lead only to a small measurement strength and, consequently, a large correlation length (due to the RG flow). While the current discussion is about hTC, a similar ballistic limit also occurs in the symmetry-class-D monitored circuit that arises from the square-lattice toric code with coherent errors\cite{Venn_2023}. 

We remark that, as indicated by Eq. \eqref{eq:RG_ClassD}, the weak coupling limit $g\rightarrow 0$ represents a stable phase in the alternative replica limit $R\rightarrow 0$. This phase corresponds to the disordered thermal metal phase in the two-dimensional class-D Anderson localization problem\cite{SenthilFisher2000,BocquetSerbanZirnbauerDirtySC2000,ChalkerKagalovskyEtAlThermalMetalPRB2001}.

Based on the analysis above, we expect that a generic phase diagram of a 1+1D class-D Majorana monitored circuit will be occupied only by distinct area-law phases separated by phase transitions. One of the area-law phases is represented by the limit with $\theta_l =0$ for all links $l$. Obviously, this area-law phase of the monitored circuit corresponds to the decodable phase of the hTC with $Z$-type coherent error. Other area-law phases correspond to undecodable phases. All of these phases have finite correlation lengths. For the decoding problem of hTC, we expect the logical error rate after the optimal decoding of $|{\cal E}_Z\rangle$ to saturate exponentially to the thermodynamical value as the size of the code increases. 

Later, we will see in Sec. \ref{sec:sTC} that class-D Majorana minotored circuits also naturally arise from the square-lattice toric code subject to coherent errors. The corresponding monitored circuit has a geometry that closely resembles the hTC counterpart shown in Fig. \ref{fig:MajCircuit_hTCZ}. For the numerical investigation of decodability in symmetry class D, we will focus on the square-lattice case as a representative example to probe the universal properties and overall structure of the phase diagram. The precise locations of the decodability transitions (i.e., the error thresholds) depend on the specific parameterization of the errors and the lattice geometry. A detailed numerical study of the decodability phase diagram for the honeycomb lattice toric code with $Z$-type coherent errors will be deferred to future work.

\section{Square-lattice Toric Code}
\label{sec:sTC}
In this section, we consider the square-lattice toric code (sTC) with coherent errors. The $X$-type coherent error $\prod_l e^{\ii \theta_l X_l}$ and the $Z$-type coherent errors $\prod_l e^{\ii \theta_l Z_l}$ are related by a symmetry in the sTC that exchanges the $X_l$ and $Z_l$ operators for all qubits. Hence, we will focus only on the $X$-type coherent error in the following. 

Coherent errors on sTC with a uniform angle, such as $\prod_l e^{\ii \theta X_l}$ has been studied in several previous works\cite{bravyi2018correcting,BaoUnitaryErrors2024,Venn_2023,BehrendsBeri2025,BehrendsSurfaceCodesQuantumCircuits}. In particular, Refs. \onlinecite{Venn_2023,BehrendsSurfaceCodesQuantumCircuits} mapped the decoding problem of sTC with uniform $X$-type error, $\prod_l e^{\ii \theta X_l}$, to a 2D Majorana scattering network model. The microscopic scattering parameters of this network model were found to match those in the scattering network models for the symmetry-class-D Anderson localization problem. This Majorana scattering network model was also treated as a limit of symmetry class DIII models in these works. Ref. \onlinecite{Venn_2023} utilized this network model to perform numerical simulations of the optimal decoder for the square-lattice surface code. Ref. \onlinecite{BehrendsSurfaceCodesQuantumCircuits} presented a similar analysis on an approximated version of the optimal decoder, which is technically sub-optimal. Both Refs. \onlinecite{Venn_2023,BehrendsSurfaceCodesQuantumCircuits} reported thresholds $\theta_{\rm th}$ that separate an decodable insulating phase of sTC (with $0\leq\theta<\theta_{\rm th}$) and a metallic undecodable phase (with $\theta_{\rm th}< \theta \leq\pi/4$). Here, the terms ``insulating" and ``metallic" are adapted from the Anderson localization problem. In the language of the dual quantum circuit, these two phases would correspond to the area-law phase and the critical phase, respectively. 

While the sub-optimal decoder is beyond the scope of our paper, the results in Ref. \onlinecite{Venn_2023} about the optimal decoder raise an interesting question when we carefully scrutinize the difference between the 2D Anderson localization problem and the decoding problem. As reviewed in Sec. \ref{sec:SymmetryClass_hTCX} and \ref{sec:SymmetryClass_hTCZ}, these two types of problems can be unified under the replica NLSM framework. A key difference is the replica limit: The Anderson localization problem is captured by the limit $R\rightarrow 0$, while the decoding problem, together with its dual monitored circuit description, is instead described by the limit $ R\rightarrow 1$. Physically, this difference arises from the difference between the random disorder probability distribution in the Anderson problem and the syndrome probability distribution in the decoding problem. The former is usually prefixed (by materials or models) without any reference to the system's state, while the latter follows from the Born rule for measuring quantum wavefunctions. As mentioned in Sec. \ref{sec:SymmetryClass_hTCX} and \ref{sec:SymmetryClass_hTCZ}, in the 2D Anderson localization problem ($R\rightarrow 0$), the metallic phase is stable under RG in both symmetry classes D and DIII. In contrast, for the decoding problem ($R\rightarrow 1$), this metallic (or critical) phase remains stable under RG only in symmetry class DIII, but not in symmetry class D. Hence, the RG result is at odds with the putative metallic phase in a symmetry-class-D decoding problem reported in Ref. \onlinecite{Venn_2023}. 

In addition to whether the metallic phase is stable, the decoding problems of symmetry classes D and DIII also exhibit different topological classification ($\mathbb{Z}$ v.s. $\mathbb{Z}_2$) for the insulating phases (which are equivalent to the area-law phases in the dual monitored circuits). Therefore, Ref. \onlinecite{Venn_2023,BehrendsSurfaceCodesQuantumCircuits}'s method that treats symmetry class D as a limit of DIII is not ideal for understanding the global structure of the decodability phase diagram.

In this section, we study the decoding problem of sTC with $X$-type errors from both the dual-monitored circuit and the error-correction perspectives. Similar to the hTC case in Sec. \ref{sec:SymmetryClass_hTCZ}, we will emphasize the essential role of TR symmetry in fixing the symmetry class to be D. Our approach provides a robust reasoning for how the symmetry classes emerge and enables the identification of the symmetry class beyond the specific form of coherent error. Furthermore, we consider a two-parameter coherent error model of the $X$-type that allows for spatially non-uniform error angles $\theta_l$. This new error model reveals an undecodable phase that is topologically distinct from the decodable phase, as well as a new decodability transition. We map out the decodability phase diagram of this two-parameter coherent error model by simulating the dual Majorana dynamics. We also revisit the putative metallic phase reported in Ref. \onlinecite{Venn_2023} using the new error model. Furthermore, we directly study the logical error rates and decodability transition of the square-lattice surface code (after optimal decoding) under this new error model.

\subsection{Stabilizers, Coherent Error Models, and Time-Reversal Symmetry}
We consider the sTC with the qubits living on the links $l$ of the square lattice. The vertex and plaquette stabilizers (see left panel of Fig. \ref{fig:squareTC}) are given by 
\begin{align}
   A_{v} = \prod_{v\in l } X_l,~~~~~~B_{p} = \prod_{l\in p} Z_l.
\end{align}
where $v$ and $p$ label the vertices and square plaquettes. 

As discussed above, we will focus on the $X$-type coherent error $\prod_l e^{\ii \theta_l X_l}$ that rotates the logical state $|\Omega\rangle$ into the error-corrupted state $\cEx$. $\cEx$ will only produce syndromes $\{s_p\}$ with respect to the plaquette stabilizers $B_p$. The probability distribution of the syndromes $\{s_p\}$ is given by
\begin{align}
     \tcP_X(\{s_p\}) = \langle{\cal E}_X |\prod_p\left(\frac{1+s_p B_p}{2}\right) |{\cal E}_X\rangle.
\end{align}
We will focus on the behavior of an infinite system where this distribution is independent of the choice of the logical state $|\Omega\rangle$. 

We introduce the ordinary anti-unitary TR symmetry 
\begin{align}
    \cT = \left(\prod_l \ii Y_l\right){\cal K} \label{eq:TR2}
\end{align}
that acts on the qubit operators as $\cT: X_l\rightarrow-X_l,~ Z_l\rightarrow-Z_l,~ Y_l\rightarrow-Y_l$. Under this TR action, both the stabilizers and the coherent error are symmetric:
\begin{align}
    {\cal T}:~~ 
        A_v\rightarrow A_v,~~B_p\rightarrow B_p,~~ e^{\ii \theta_l X_l} \rightarrow e^{\ii \theta_l X_l}.
        \label{eq:TR_Error2}
\end{align}
Therefore, the error-corrupted state $\cEx$ is TR symmetric.

\subsection{Statistical Model, Majorana Monitored Circuit, and Symmetry Class}
\begin{figure}
    \centering
    \includegraphics[width=0.9\linewidth]{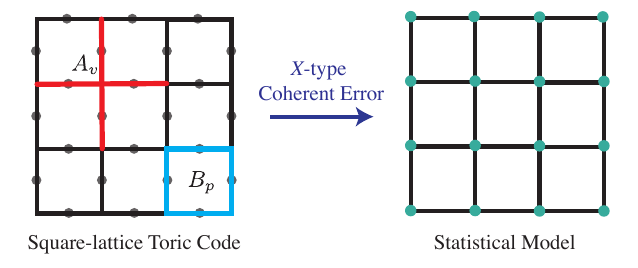}
    \caption{Square-lattice toric code with the qubits (black dots) on the links of the lattice (left panel). The vertex stabilizer $A_v$ (red) acts on the four links connected to the vertex $v$. The plaquette stabilizer $B_p$ (blue) acts on the four links of the plaquette $p$. The syndrome distribution generated by the $X$-type coherent error can be calculated from a disordered statistical model with Ising spins (green dots) on the square lattice.
    }
    \label{fig:squareTC}
\end{figure}

The syndrome distribution $\tcP_X(\{s_p\})$ in $|{\cal E}_X\rangle$ can be written in terms of the partition function $\tcZ_X(\{s_p\})$ of a classical disordered statistical model on the square lattice with classical Ising spins $\tau_v =\pm 1$ living on the vertices (see the right panel of Fig. \ref{fig:squareTC}):
\begin{equation}
     \tcP_X(\{s_p\}) = |\tcZ_{X,\{s_p\}}|^2,
\end{equation}
where the partition function $\tcZ_{X,\{s_p\}}$ is given by 
\begin{align}
    \tcZ_{X,\{s_p\}} = {\cal N} \sum_{\{\tau_v =\pm1\}} \exp{\left[ - \sum_{\langle v,v'\rangle} J_{v,v'} \eta_{v,v'} \tau_v \tau_{v'} \right]}.
    \label{eq:sTCX_DisorderedPartitionFunction}
\end{align}
Here, $J_{v,v'}$ is a complex nearest-neighbor Ising coupling satisfying $e^{2J_{v,v'}} = \ii \tan\theta_l$ for every pair of neighboring sites $v$ and $v'$
connected by the link $l=\langle v,v'\rangle$. The variable $\eta_{v,v'} = \pm 1$ is a random-bond disorder on the square lattice related to the syndromes via $\prod_{l\in p} \eta_l = s_p$. As in previous cases, different $\{\eta_l\}$ configurations associated with the same syndrome are gauge-equivalent and yield the same partition function. Naturally, $\tcZ_{X,\{s_p\}}$ can be relabeled as $\tcZ_{X,\{\eta_l\}} $, which highlights
its dependence on the random bond configuration $\{\eta_l\}$. The disorder-averaged free energy $\tilde{F}_X$ that is tied to the codes' intrinsic syndrome distribution and decodability is given by
\begin{align}
   \tilde{F}_X \propto  &\sum_{\{\eta_l\}} -|\tcZ_{X,\{\eta_l\}}|^2\log |\tcZ_{X,\{\eta_l\}}|^2.
\end{align}

\begin{figure}
    \centering
    \includegraphics[width=0.95\linewidth]{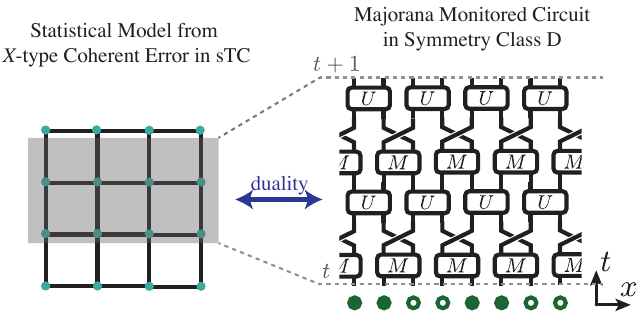}
    \caption{The statistical model (left panel) that describes the sydrome distribution of the square-lattice toric code with $X$-type coherent error can be dualized to a 1+1D monitored circuit (right panel) acting a Majorana chain. The Majorana chain has two sublattices $A$ (solid green dots) and $B$ (hollow green dots). This Majorana monitored circuit is non-interacting and belongs to AZ symmetry class D. The shaded part of the statistical model is mapped to the circuit block between times $t$ and $t+1$. Unitary gates and measurements are labeled by $U$ and $M$, respectively.}
    \label{fig:MajCircuit_sTC}
\end{figure}

Similar to previous cases, we can map this disordered statistical model to a dual non-interacting monitored quantum circuit acting on a chain of Majorana modes $\psi_n$ (see Fig. \ref{fig:MajCircuit_sTC}). The shaded part of the statistical model is mapped to the circuit block between times $t$ and $t+1$. In particular, every vertical link in the statistical model is mapped to a random unitary gate $U_{\eta_l}$ as defined in Eq. \eqref{eq:RandUnitary}. Every horizontal link is mapped to a generalized measurement ${\cal M}_l$ as defined in Eq. \eqref{eq:KrausOp}. Naively, the circuit has a two-site unit cell in the spatial direction. However, for our later discussion of TR symmetry, we introduce a four-site unit cell with two sublattices $A$ and $B$ (depicted as solid and hollow green dots in the right panel of Fig. \ref{fig:MajCircuit_sTC}). The circuit block shown in Fig. \ref{fig:MajCircuit_sTC} repeats every unit time step. 

Now, we introduce the TR symmetry $\cT_c$ of this monitored circuit, which acts on the Majorana modes as  
\begin{align}
    \cT_c: \begin{cases}
        \psi_n \rightarrow \psi_n,~~& n\in \text{sublattice}~A, \\
        \psi_n \rightarrow -\psi_n,~~& n\in \text{sublattice}~B.
    \end{cases}
\end{align}
This TR symmetry acts on the Majorana monitored circuit in a similar way as in the case of hTC with $Z$-type coherent error (see Sec. \ref{sec:hTCZ}). Most importantly, the Majorana monitored circuit in the current case also commutes with the TR action $\cT_c$. Since $\cT_c^2=1$, the Majorana monitored dynamics dual to the decoding problem of sTC subject to $X$-type coherent errors belong to symmetry class D\cite{bhuiyan2025free}. This statement holds for any spatial distribution of the error angles $\{\theta_l\}$. 

We remark that this method, which identifies the symmetry class of the decoding problem via the symmetry of the dual monitored circuit, is applicable to more general coherent errors, in which case the monitored circuit can become interacting. For example, the decoding problem remains in symmetry class D under the coherent error $\prod_l e^{\ii (\alpha_lX_l + \beta_l Y_l+\gamma_l Z_l)}$. Correlated coherent error can break the TR symmetry $\cT$.

\subsection{Two-parameter Coherent Error Model and Solvable Limits}
\label{sec:two-parameter_sTC}
\begin{figure*}
    \centering
    \includegraphics[width=0.95\linewidth]{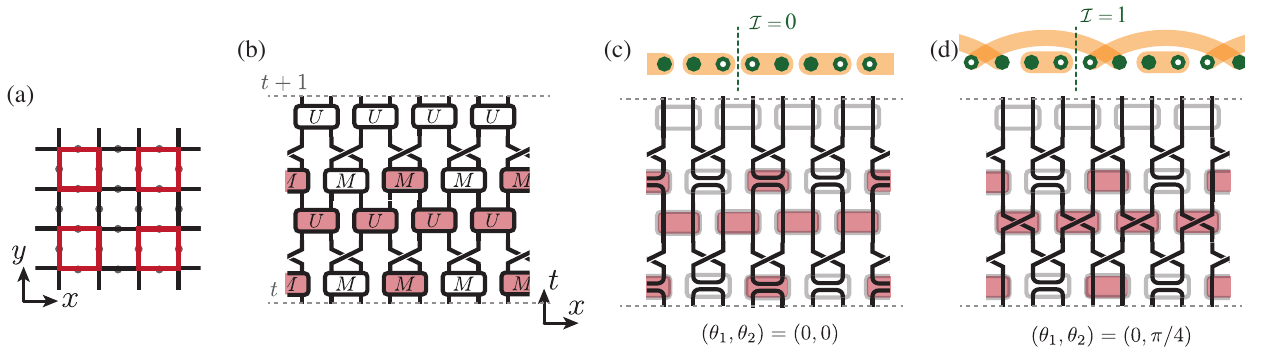}
    \caption{(a) The two-parameter coherent error model: $e^{\ii\theta_1 X}$ on the black links and $e^{\ii\theta_2 X}$ on the red links. (b) The Marjoana monitored circuit model associated with the two-parameter coherent error model. The red gates are associated with red links on the square lattice, and the white gates are associated with black links. The pattern of gates in the monitored circuit repeats in every unit time step. (c) The Majorana monitored circuit with $(\theta_1,\theta_2)= (0,0)$ stabilizes a nearest-neighbor Majorana dimerization pattern (shown at the top) with topological index ${\cal I}=0$. The topological index is defined with respect to a cut (green dashed line) on the Majorana chain.
    (d) The Majorana monitored circuit with $(\theta_1,\theta_2)= (0,\pi/4)$ stabilizes a Majorana dimerization pattern (shown at the top) with topological index ${\cal I}=1$. }
    \label{fig:TwoParameter_sTC}
\end{figure*}

In the following, we introduce a two-parameter family of $X$-type coherent errors following the pattern shown in Fig. \ref{fig:TwoParameter_sTC}(a): For the qubits on the black links in the sTC, the error angles are $\theta_l = \theta_1$; for the qubits on the red links, the error angles are $\theta_l=\theta_2$. When dualized to the Majorana monitored circuit, the red links give rise to the red gates, while the black links correspond to the white gates as shown in Fig. \ref{fig:TwoParameter_sTC}(b).

First, let's briefly discuss the motivation for studying the two-parameter error model in this case. Note that the special line with $\theta_1 =\theta_2 =\theta$ in our parameter space corresponds to the uniform coherent error model studied in previous works\cite{bravyi2018correcting,BaoUnitaryErrors2024,Venn_2023,BehrendsBeri2025,BehrendsSurfaceCodesQuantumCircuits}. Along this special line, in the language of monitored circuits, only one of the $\mathbb{Z}$-classified area-law phases was found. This phase corresponds to the decodable phase of the sTC. The tension between the RG-based expectation and the numerical observation\cite{Venn_2023} of an undecodable metallic (or critical) phase raises interesting questions regarding the nature of the reported decodability transition at $\theta=\theta_{\rm th}$. Now, allowing the parameters $\theta_1$ and $\theta_2$ to vary independently, we can more comprehensively explore the decodability phase diagram over an expanded regime. As we show below, the new decodability transition between two topologically distinct area-law phases will appear in the expanded phase diagram. This transition is generic, even if we introduce further spatial variation of $\theta_l$. But the location of this transition, i.e., the associated error threshold, can depend on the microscopic details of the error model. 

Next, we discuss the solvable limits of the Majorana monitored circuit. These limits will help us understand the structure of the decodability phase diagram. Obviously, $(\theta_1,\theta_2) = (0,0)$ is a solvable point that represents the decodable phase. The corresponding monitored circuit stabilizes the Majorana dimerization pattern shown in Fig. \ref{fig:TwoParameter_sTC}(c). Using the criteria discussed in Sec. \ref{sec:SymmetryClass_hTCZ}, the corresponding integer-valued topological index $\cal I$ defined with respect to the cut [green dashed line in Fig. \ref{fig:TwoParameter_sTC}(c)] is ${\cal I} = 0$.

The solvable point $(\theta_1,\theta_2) = (0,\pi/4)$ represents a different area-law phase of the monitored circuit where the Majorana modes dimerize according to Fig. \ref{fig:MajCircuit_hTCZ}(d). The corresponding topological index is ${\cal I} = 1$. The point $(\theta_1,\theta_2) = (\pi/4,0)$ is related to $(0,\pi/4)$ by a translation of the entire sTC by one site in both the $x$- and $y$-directions. This point also represents an area-law phase with topological index ${\cal I} = 1$. 

The limit $(\theta_1,\theta_2) = (\pi/4,\pi/4)$ is the special point where the entire Majorana monitored circuit is essentially composed of unitary SWAP gates. The Majorana modes evolve ballistically in this limit. As discussed in Sec. \ref{sec:SymmetryClass_hTCZ}, a small deviation from this limit may result in a large correlation length as a consequence of the RG flow in the replica limit $R\rightarrow 1$. The previously reported metallic phase\cite{BehrendsSurfaceCodesQuantumCircuits,Venn_2023}, identified via numerical simulations, was located in the neighborhood of this limit $(\theta_1,\theta_2) = (\pi/4,\pi/4)$.

Next, we discuss the relationship between different values of $(\theta_1, \theta_2)$. First of all, coherent errors with $(\theta_1 + n_1 \pi/2, \theta_2 + n_2 \pi/2)$ for all $n_{1,2}\in \mathbb{Z}$ give rise to same error-corrupted state $\cEx$. That is because the difference between these $(\theta_1, \theta_2)$ parameters amounts to additional products of vertex stabilizers $A_v$. These additional stabilizers do not affect the initial logical state $|\Omega\rangle$. Therefore, coherent errors with $(\theta_1 + n_1 \pi/2, \theta_2 + n_2 \pi/2)$ lead to the same decodability. Secondly, $(\theta_1, \theta_2)$ can be mapped to $(\theta_2, \theta_1)$ after a spatial transition of sTC by one site in both the $x$- and $y$-directions. Further more, the $X$-type coherent errors with $( \theta_1,\theta_2)$, $( -\theta_1,\theta_2)$, $( \theta_1,-\theta_2)$ and $(-\theta_1, -\theta_2)$ share the same decodablity. For example, the error-corrupted states with $( \theta_1,\theta_2)$ and $( -\theta_1,\theta_2)$ are related by a product of plaquette stabilizers acting only on the black links (see Fig. \ref{fig:TwoParameter_sTC}(a)). Since the syndromes $\{s_p\}$ to be decoded in the error-corrupted state $\cEx$ are extracted from measuring the plaquette stabilizers $B_p$, additional multiplications of $\cEx$ by $B_p$'s do not change the state's decodability. With all these relations between different values of $(\theta_1,\theta_2)$, the full 2-parameter decodability phase diagram can be represented by the fundamental domain, which we choose to be the regime with $0\leq \theta_1 \leq \theta_2 \leq \pi/4$. 

Lastly, we discuss the translation of the sTC by one site in the $x$-direction, which corresponds to a translation $T_2$ of the Majorana chain in the dual monitored circuit by two Majorana sites. This $T_2$ operation interchanges the $A$ and $B$ sublattices on the Majorana chain. Consequently, an area-law phase with topological index $\cal I$ transforms into an area-law phase with topological index $-\cal I$. More interestingly, a $T_2$-invariant area-law phase must have $\cI = 0$, which is the same as the decodable phase. Here, we are assuming that the cut for defining $\cal I$ is between two $B$ sites (or between two $A$ sites) as shown in Fig. \ref{fig:TwoParameter_sTC}. In our two-parameter error model, a generic $\twotheta$ breaks the $T_2$ symmetry explicitly, except for the case with $\theta_1 = \theta_2$, i.e., the case of uniform coherent error. Therefore, the area-law phase along the line $\theta_1 = \theta_2$ should be the decodable phase with $\cI = 0$. As discussed in Sec. \ref{sec:SymmetryClass_hTCZ}, it is likely that a generic phase diagram of a class-D monitored circuit contains only area-law phases and their transitions. A possible scenario for the case of the uniform error, namely $\theta_1=\theta_2$, is that the decodable area-law phase with $\cI = 0$ occupies the entire range $0\leq \theta_1 = \theta_2< \pi/4$. Recall that the point with $\theta_1 = \theta_2= \pi/4$ is the special point where the circuit becomes unitary.

\subsection{Phase Diagram }
\label{sec:sTC_PhaseDiagram}

\begin{figure*}[t!]
    \centering
    \includegraphics[width=0.85\linewidth]{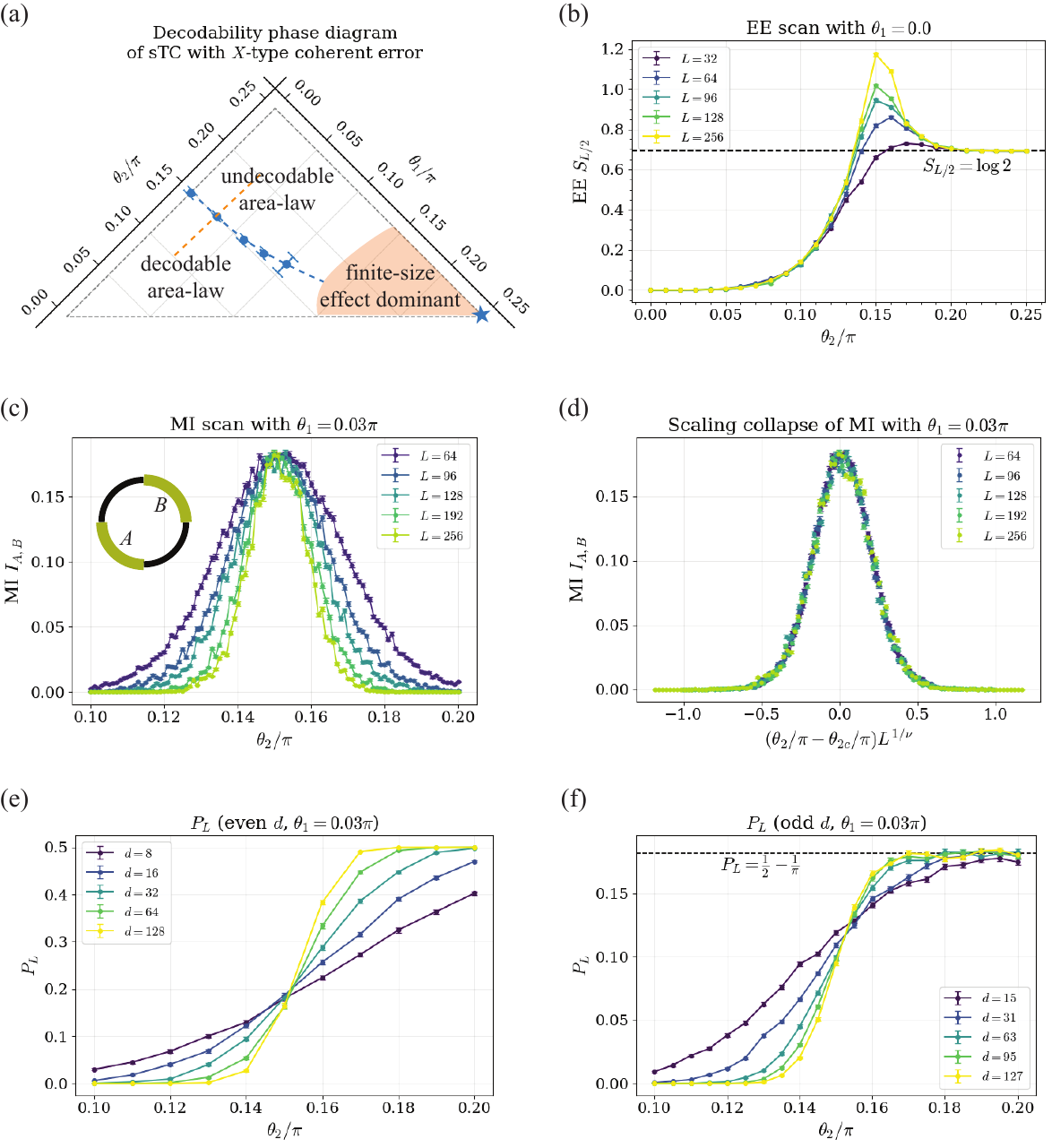}
    \caption{(a) Phase diagram of the two-parameter coherent error model on the square lattice. We find a decodable area-law phase ($\cI=0$) and an undecodable area-law phase ($\cI=1$) separated by a decodability phase transition (blue dashed line). $\theta_1=\theta_2=\pi/4$ (star) is a special point with volume-law EE scaling. Around this point is a region with large correlation lengths (orange shade), where our numerics are dominated by finite-size effects. 
    (b) Half-system EE $S_{L/2}$ along the line $\theta_1=0$. 
    (c) A representative scan of mutual information (MI) along a 1D slice of the phase diagram (orange dashed line in (a)). The MI $I_{A,B}$ is calculated between two antipodal quarter-subsystems $A$ and $B$ (inset). Phase boundary (blue dots in (a)) and the onset of strong finite-size effects (App. \ref{app:finitesize}) are obtained from the MI data.    
    (d) The scaling collapse of the MI data in (c).
    (e)(f) Logical error rate $P_L$ [Eq. \eqref{eq:ler}] for the rotated surface code with even and odd code distances along a 1D slice of the phase diagram (orange dashed line in (a)). The crossing points of $P_L$ near $\theta_2\approx 0.15\pi$ agree with the critical value extracted from the scaling collapse in (d).
    The MI data are averaged over 5000-10000 samples, $P_L$ for even $d$ over 1000 samples, $P_L$ for odd $d$ over 5000 samples, and EE over 1000 samples. }
    \label{fig:squarephase}
\end{figure*}

The phase diagram of the two-parameter model of sTC with coherent errors is presented in Fig. \ref{fig:squarephase}(a). Only the fundamental domain $0\leq\theta_1\leq\theta_2\leq\pi/4$ is shown. Analogous to the hTC case, the phase diagram is obtained by numerically simulating the non-interacting monitored circuit shown in Fig. \ref{fig:TwoParameter_sTC}(b) on periodic Majorana chains of various lengths $L$. The spacetime aspect ratio is fixed. The numerical simulation is done with fermionic Gaussian circuit methods (see App. \ref{app:algocircuit}). The different phases are distinguished by entanglement properties of the final states of the circuit, averaged over quantum trajectories. In addition to simulating the dual Majoana circuit, we also directly study the decoding problem on a square-lattice surface code using the two-parameter model to corroborate our results on the decodability transitions.

The phase diagram Fig.\ref{fig:squarephase}(a) contains a decodable area-law phase and an undecodable area-law phase. Here ``decodability" refers to the corresponding error-corrupted sTC, while ``area-law" refers to the entanglement scaling of the circuit final state. Fig.\ref{fig:squarephase}(b) shows that inside each area-law phase, the half-system EE $S_{L/2}$ quickly saturates to an ${\cal O}(1)$ value as $L$ increases. The decodable area-law phase grows out of the solvable point $\twotheta = (0,0)$ shown in Fig.\ref{fig:TwoParameter_sTC}(c) with topological index $\cI =0$ and the undecodable area-law phase contains the solvable point $\twotheta = (0,\pi/4)$  [shown in Fig.\ref{fig:TwoParameter_sTC}(d)] with topological index $\cI=1$. The two area-law phases are separated by a direct phase transition without any intervening critical phase, in contrast to the hTC case [Fig.\ref{fig:honeyphase}(a)]. This is consistent with the observation that a critical phase should be absent in symmetry class D, as discussed in Sec. \ref{sec:SymmetryClass_hTCZ}.

The special point at $(\theta_1, \theta_2) = (\pi/4, \pi/4)$ exhibits volume-law EE, which we confirm in our numerics. This point also corresponds to having ``maximal error" on each qubit of sTC. The corresponding Majorana circuit at this point consists of only unitary SWAP gates, and the Majorana fermions propagate ballistically. Around this point, we encounter a region where the finite-size effect dominates in our numerics (Fig. \ref{fig:squarephase}(a) orange colored region). This observation is potentially the manifestation of the large correlation length $\xi$ expected in the neighborhood of $(\theta_1, \theta_2) = (\pi/4, \pi/4)$, which we argued in Sec. \ref{sec:SymmetryClass_hTCZ} using RG. In App. \ref{app:finitesize}, we present numerical data that are consistent with this expectation. For example, the average second moment of the fermion correlation $\overline{G}(r) \equiv \overline{|\braket{\ii \psi_n\psi_{n+r}}|^2}$ in the Majorana circuit should exhibit a power-law behavior in a true critical phase. We numerically confirm this behavior in Fig. \ref{fig:twopointcorr}(a) for the critical phase of hTC with $X$-type error. In comparison, as we can see from Fig. \ref{fig:twopointcorr}(b), the $\overline{G}(r)$ does not fit well with a simple power law in the orange region of the phase diagram Fig. \ref{fig:squarephase}(a). Also, in App. \ref{app:finitesize}, we attempt to locate the possible phase boundary between the decodable phase and the would-be critical phase in the orange region. We observe a system-size-dependent drift of the numerically extracted phase boundary (via MI crossing), which is consistent with our RG analysis.

Note that Ref. \onlinecite{Venn_2023} numerically studied the case of uniform coherent error, namely $\theta_1 =\theta_2$, and observed entanglement and transport properties that resemble a critical (or metallic) phase with logarithmic EE scaling in the orange region of Fig. \ref{fig:squarephase}(a). However, we've discussed the instability of the weak-coupling critical phase in class-D monitored circuits in Sec. \ref{sec:SymmetryClass_hTCZ}. Also, we found large finite-size effects in our numerical results within this region, as mentioned above (see App. \ref{app:finitesize}). Furthermore, we've shown that the only possible area-law phase with $\theta_1 =\theta_2$ should have the topological index $\cI=0$, which is the same as the decodable phase. It is possible that the whole range with $0\leq \theta_1=\theta_2<\pi/4$ is decodable, and the critical-phase-like behavior reported in Ref. \onlinecite{Venn_2023} is due to correlation lengths being larger than system sizes in the numerical simulation. 

We remark that, for the sTC with uniform coherent error, the reports of the critical phase and error threshold in the context of suboptimal decoders (or approximated optimal decoders) \cite{bravyi2018correcting,BehrendsSurfaceCodesQuantumCircuits} do not contradict our conclusion, as we focus on the code's decodability under the optimal decoding scheme.

To find the phase boundary between the two area-law phases, we calculate the MI $I_{A,B}$ of two antipodal quarter-subsystems $A$ and $B$ (see inset of Fig. \ref{fig:squarephase}(c)) in the circuit's final states (averaged over different quantum trajectories). Inside each area-law phase, the final states are short-range correlated. $I_{A,B}$, which measures the correlation between the two disjoint subsystems $A$ and $B$, should approach $0$ exponentially with system size $L$. On the other hand, the emergent conformal symmetry at the phase transition should lead to a system-size independent MI $I_{A,B}$\cite{LiConformal2021}. The phase boundary between the decodable and undecodable phases in Fig. \ref{fig:squarephase}(a) is located by finding $\twotheta$'s where $I_{A,B}$ becomes system size independent along several slices of the parameter space (similar to finding MI crossings in the hTC case).

An example of these MI behaviors is shown in Fig. \ref{fig:squarephase}(c), which plots the MI $I_{A,B}$ along a slice of the phase diagram with $\theta_1 = 0.03\pi$. The scaling collapse of the data in Fig. \ref{fig:squarephase}(c) is shown in Fig. \ref{fig:squarephase}(d). The fitted critical value is $\theta_{2c}/\pi = 0.1504^{+0.0007}_{-0.0012}$. The correlation length exponent is $\nu=1.75^{+0.12}_{-0.06}$. Following our discussion in Sec. \ref{sec:SymmetryClass_hTCZ}, this phase transition should be captured by the continuum NLSM with target manifold $\SO(2R)/\U(R)$ in the limit $R\rightarrow 1$. Interestingly, 
Ref.~\onlinecite{GYZhu2025SelfDual} found the 
same universality class and the same NLSM appearing in the microscopically unrelated context of measurement- and decoherence-induced self-dual criticality of the 2D toric code, and reported an exponent $\nu$ that is consistent with our result.

Next, we investigate the decodability phase diagram directly from the perspective of error correction. In sTC, the fundamental decodability can be measured by how well the optimal decoder (with Pauli correction) can recover the quantum information in the initial state $\ket{\Omega}$. On the error-corrupted state $\ket{\cE_X}$, syndrome measurements project it to an (unnormalized) post-measurement state 
$\prod_p\left(\frac{1+s_p B_p}{2}\right)\ket{\cE_X}$
when the measured syndromes are $\{s_p\}$. The decoder chooses an error-correcting Pauli string $\cC_s$ according to the syndrome $s_p$ and applies this correction to the post-measurement state, resulting in a corrected state $\ket{\psi_s}$ that is free of syndrome:
\begin{equation}
    \ket{\psi_s} =\mathcal{N} \cC_s\prod_p\left(\frac{1+s_p B_p}{2}\right)\ket{\cE_X},
\end{equation}
where $\mathcal{N}$ is a constant of normalization. The logical error rate $P_L$ is defined via the the overlap between this final state $\ket{\psi_s}$ and the initial logical state $|\Omega\rangle$,
\begin{equation}
    1-P_L \equiv |\braket{\psi_s|\Omega}|^2. \label{eq:ler} 
\end{equation}
The performance of a decoder is then evaluated using the average logical error rate $P_L$ over all syndromes.

As discussed, to understand the code's fundamental decodability, one should consider the performance of the optimal decoder. Specifically, we will consider the maximal likelihood (ML) decoder, which is optimal and always chooses a correction $\cC_s$ that minimizes the logical error rate $P_L$ for every syndrome $\{s_p\}$\cite{Dennis2002topo}. In the decodable phase, $P_L$ should vanish as the code distance $d\to \infty$, whereas in the undecodable phase, $P_L$ should approach a finite positive number as $d\to \infty$.

In the following, we numerically study the behavior of the error rate $P_L$ averaged over different syndromes $\{s_p\}$ using the ML decoder. To compute $P_L$ in the sTC and make connections with previous works, we consider the geometry of the rotated surface code (rSC) \cite{Wen2003quantum,Bombin2007optimal,bravyi2018correcting, GoogleBelowThreshold_s}, which is essentially the sTC with a particular choice of open boundary conditions such that it stores one logical qubit. We denote the code distance by $d$. See App. \ref{app:algosample} for more details.  

It was noted \cite{bravyi2018correcting,Cheng_2025} that rSC under $X$-type coherent error behaves very differently for even and odd code distances $d$. For odd $d$, the state after syndrome measurement and Pauli correction is, 
\begin{equation}
    \ket{\psi_s} = e^{\ii \varphi_s \overline{X}}\ket{
    \Omega},
    \label{eq:coherent_LogicalError}
\end{equation}
where $\overline{X}$ is the logical $X$ operator and $\varphi_s$ depends only on the syndrome. This is a unitary rotation in the logical space.  For even $d$, the syndrome measurement and correction is effectively a weak measurement with post-selection on the logical space,
\begin{equation}
    \ket{\psi_s} \propto (1+\beta_s \overline{X})\ket{\Omega},
    \label{eq:incoherent_LogicalError}
\end{equation}
where the strength of weak measurement $\beta_s$ depends only on the syndrome $s$. Heuristically, the worst possible logical error of the form Eq. \eqref{eq:coherent_LogicalError} would have $\theta_s$ appearing as a random angle for different syndromes $\{s_p\}$. 
The worst-case logical error of the form in Eq.~\eqref{eq:incoherent_LogicalError} occurs when $\beta_s=\pm1$ and the protocol effectively becomes a projective $\overline{X}$ measurement.

To perform a concrete calculation of the average $P_L$, we fix the initial state to be the logical $\overline{Z}$ eigenstate $\ket{\Omega} = \ket{\overline{0}} $. The syndromes $\{s_p\}$ are sampled using the algorithm by Ref.  \onlinecite{bravyi2018correcting}, which maps the error-corrupted rSC to a fermionic Gaussian tensor network (see App. \ref{app:algosample} for a brief review). The ML decoder is then used to find $\ket{\psi_s}$. The logical error rate $P_L$ is calculated using Eq. \eqref{eq:ler}, averaged over many syndrome samples. Figs. \ref{fig:squarephase}(e) and (f) show $P_L$ for even and odd code distances $d$ for fixed $\theta_1 = 0.03\pi$ and varying $\theta_2$ [along the orange dashed line in Fig. \ref{fig:squarephase}(a)]. The decodability transition at $\theta_{2c}\approx 0.15\pi$ identified using MI [Fig. \ref{fig:squarephase}(d)] appears as the crossings in the $P_L$ scans for different code distances $d$. For $\theta_{2}<0.15 \pi$, the code is in the decodable phase, and $P_L$ decreases to $0$ exponentially with $d$.  For $\theta_2>0.15\pi$, the error rate $P_L$ converges to some positive values $P_L^{(\infty)}$ as $d\to \infty$, signaling the breakdown of decodability. 

Based on the heuristic picture for the worst-case logical error for even and odd code distances, in the limit of abundant errors, one expects \cite{Venn2020error,Venn_2023},
\begin{equation}
     P_L^{(\infty)}=
    \begin{cases}
    \frac{1}{2} - \frac{1}{\pi}, & d \text{ odd},\\
    \frac{1}{2}, & d \text{ even}.
    \end{cases} \label{eq:PLlimits}
\end{equation}
To provide some intuition, we briefly discuss how these heuristic values arise. The result here is natural for even $d$. To derive the result for odd $d$, we consider a random logical rotation angle $\varphi_s$ in Eq. \eqref{eq:coherent_LogicalError}, i.e., drawn from a uniform distribution in $[0,2\pi)$. The logical error rate would be $P_L = 1-|\braket{\overline{0}|\psi_s}|^2 = (\sin\vphi_s)^2$, naively. However, a decoder might choose a different error class, resulting in a final state $\overline{X}\ket{\psi_s}$    
and a logical error rate $ P_L' = 1-|\braket{\overline{0}|\overline{X}|\psi_s}|^2 = (\cos\vphi_s)^2$.
The optimal decoder should choose the error class between the two that minimizes the logical error rate:
\begin{equation}
    P_L^{\rm (op)} = \min\{(\sin\vphi_s)^2, (\cos\vphi_s)^2\}.
\end{equation}
Therefore, the average logical error rate for the optimal decoder deep inside the undecodable phase should be
\begin{equation}
    P_L^{(\infty)} = \frac{1}{2\pi}\int_0^{2\pi} {\rm d}\theta \min\{(\sin\vphi_s)^2, (\cos\vphi_s)^2\} = \frac{1}{2} - \frac{1}{\pi}.
\end{equation}
The heuristic values in Eq. \eqref{eq:PLlimits} agree with the large $\theta_2$ limits of Figs. \ref{fig:squarephase}(e) and (f) for large system sizes $L$, where the code is deep inside the undecodable phase.

Now, we conclude that the decodability transition obtained directly by investigating the post-optimal-decoding logical error rate in the rSC corroborates our results from studying the class-D Majorana monitored circuit dual to this decoding problem. We reiterate that this decodability transition occurs in the regime with $\theta_1\neq \theta_2$, i.e., non-uniform coherent error. From the phase diagram Fig. \ref{fig:squarephase}(a), we see that the decodability of sTC or rSC is more vulnerable to coherent errors with $\theta_1\neq \theta_2$ than the uniform ones.

\section{Conclusions and Discussion}
In this paper, we studied the decodability of the square-lattice toric code (sTC) and the honeycomb-lattice toric code (hTC) subject to coherent errors. We focused on the $X$-type and $Z$-type single-qubit coherent error generated by the unitary rotation of the physical qubits about the $X$- and $Z$-axis. We established a duality between the optimal Paui decoding problem in these error-corrupted toric codes and 1+1D monitored dynamics of non-interacting Majorana fermions. This duality shows that the AZ symmetry classes of the dual Majorana dynamics govern the universal structure of the corresponding decodability phase diagram. In particular, two symmetry classes, DIII and D, emerge from the error models and geometries we considered. From the symmetry perspective, whether the codes and error models preserve TR symmetry determines which class emerges. 

The hTC with $X$-type coherent error explicitly breaks the TR symmetry and is dual to the 1+1D Majorana monitored dynamics of symmetry class DIII. In this case, the non-interacting monitored dynamics generically admits three different phases\cite{JianRTN2022,FavaNahum2023,kells2023topological,jian2023measurement,Negari2024}, including two topologically distinct area-law phases (classified by a $\mathbb{Z}_2$ index) and a critical phase with logarithmic entanglement scaling. We showed that one of the area-law phases is dual to the decodable phase of the hTC, while the other dynamical phases correspond to undecodable phases in the hTC with $X$-type coherent error. Generic phase transitions in class DIII are expected to happen between the critical phase and an area-law phase. These phases and their transitions admit a continuum description based on the replica NLSM Eq. \eqref{eq:NLSMhTC} with $\SO(R)$ target manifold in the replica limit $R\rightarrow 1$\cite{FavaNahum2023,jian2023measurement}. The $\SO(R)$ target manifold is fully dictated by the symmetry class DIII.

We introduced a minimal two-parameter model for the $X$-type coherent error in hTC, enabling us to investigate all three phases in the symmetry class DIII and their transitions within a single microscopic model. The two-parameter model is parameterized by two error angles $\theta_{1,2}$ for the two subsets of qubits in hTC (see Fig. \ref{fig:TwoParameter_hTCX}). We study the decodability phase diagram of hTC based on the two-parameter error model using analytical and numerical methods. We identified several solvable limits of the error model (Sec. \ref{sec:two-parameter_hTCX}) and mapped out the phase diagram [Fig. \ref{fig:honeyphase}(a)] by numerically simulating the dual Majorana monitored dynamics. As expected, the generic transition out of the decodable area-law phase is dual to the area-law-to-critical measurement-induced phase transition in the monitored dynamics. 

In the numerical simulation of the Majorana monitored dynamics dual to the decoding problem of hTC, we used the scaling collapse of the MI $I_{A,B}$ between antipodal intervals $A$ and $B$ on the Majorana chain to identify the phase boundary and extracted the correlation length exponent $\nu$ [Fig. \ref{fig:honeyphase}(e)]. Our extracted $\nu$ is consistent with the previous study \cite{jian2023measurement} on the area-law-to-critical phase transition in class-DIII monitored circuits with entirely different microscopic details.

Interestingly, the two-parameter $X$-type error model for hTC contains a special line $\theta_1 =0$ where the dual Majorana monitored dynamics exhibits an extra emergent TR symmetry, which changes its symmetry class to D. Symmetry class D does not admit a stable critical phase in 1+1D (see Sec. \ref{sec:SymmetryClass_hTCZ}). Therefore, the decodability transition along this line becomes one that is dual to the direct transition between two topologically distinct area-law phases [see Fig. \ref{fig:honeyphase}(a)].

The hTC with $Z$-type coherent errors and the sTC with both types of coherent errors respect TR symmetry for generic error angles. The dual Majorana monitored dynamics inherit the TR symmetry and belong to symmetry class D. 1+1D class-D non-interacting monitored dynamics admit a $\mathbb{Z}$-classified set of area-law phases\cite{Pan_2025,XiaoKawabata2024,bhuiyan2025free,ClassificTopInsPRB2008,KitaevPeriodTable_2009,SchnyderRyu2010NJP} (labeled by a topological index $\cI\in \mathbb{Z}$). The decodable phase is dual to the area-law phase with $\cI = 0$. These area-law phases can be captured using a continuum NLSM model with target manifold $\SO(2R)/\U(R)$ in the replica limit $R\rightarrow 1$. This target manifold $\SO(2R)/\U(R)$ is dictated by the symmetry class D\cite{ChalkerKagalovskyEtAlThermalMetalPRB2001,SenthilFisher2000,IlyaReadLudwig2001,ReadLudwig2000,BocquetSerbanZirnbauerDirtySC2000,Fendley2001}. Our RG analysis on this NLSM model indicates the instability of the (weak-coupling) critical phase in the limit $R\rightarrow1$ (see Sec. \ref{sec:SymmetryClass_hTCZ}). However, the correlation length $\xi \sim a e^{6/g_0^2}$ can be large if the bare (coarse-grained) measurement strength $g_0$ in the monitored dynamics is small. Based on the reasoning above, a generic phase diagram in this symmetry class should only contain area-law phases and direct phase transitions between them. 

We remark that the replica limit $R\rightarrow 1$ captures the physics of the decoding problem (under the optimal Pauli decoding scheme) and the dual Majorana monitored circuit. The alternative replica limit $R\rightarrow 0$ corresponds to the 2D Anderson localization problem. In the class-D localization problem, the disordered thermal metal phase, which is the analog of the would-be critical phase, is stable under RG in the replica limit $R\rightarrow 0$.  

For this symmetry class D, we take the sTC subject to $X$-type coherent errors as a representative example for a more microscopic investigation. Similar to the hTC case, we introduce a minimal two-parameter model for the $X$-type coherent error that allows for spatial variations in the error angles, parameterized by $\twotheta$. The uniform coherent error model studied previously\cite{bravyi2018correcting,Venn_2023,BehrendsSurfaceCodesQuantumCircuits,BaoUnitaryErrors2024} corresponds to the case with $\theta_1 = \theta_2$ in our two-parameter model. We studied this two-parameter model both analytically and numerically. 

Previously, there were reports on the numerical evidence for a critical phase and/or a decodability transition in the uniform error model of sTC within the range $0<\theta_1 = \theta_2< \pi/4$ under optimal\cite{Venn_2023} and suboptimal decoding\cite{bravyi2018correcting,BehrendsSurfaceCodesQuantumCircuits}. While suboptimal decoding is beyond the scope of this work, a critical phase under an optimal decoding scheme in symmetry class D is at odds with the RG analysis on its instability. In our numerical study, we found that the previously reported critical phase is in a regime with large finite-size effects [see Fig. \ref{fig:squarephase}(a)]. As shown in App. \ref{app:finitesize}, in this regime, the averaged second moment of the fermion correlation function in the dual Majorana dynamics does not fit well with the power-law behavior expected for a critical phase. Our attempt to locate the transition between the decodable area-law phase and the putative critical phase found a drifting phase boundary as the system size changes. From an analytical perspective, dual Majorana dynamics in this regime have weak measurement strengths, which vanish at the special point $\theta_1 = \theta_2 =\pi/4$ where the dynamics becomes unitary. Our RG analysis suggested that this regime is subject to a large correlation length. In addition, we showed that the translation symmetry of the uniform error model ($\theta_1 =\theta_2$) pins the topological index $\cI = 0$ for the dual area-law dynamical phase (see Sec. \ref{sec:two-parameter_sTC}). Hence, it is possible that the entire line with $0<\theta_1 =\theta_2<\pi/4$ is occupied by the decodable area-law phase ($\cI=0$) under the optimal Pauli decoding scheme.

Moving away from the uniform case, our study of the two-parameter error model on the sTC further reveals the decodability transition that corresponds to the area-law-to-area-law measurement-induced phase transitions in the dual Majorana dynamics [Fig. \ref{fig:squarephase}(a)]. This result shows that the decodability of the sTC is more vulnerable to spatially varying (or oscillating) coherent error than the uniform ones, due to interference effects. The phase transition points in Fig. \ref{fig:squarephase}(a) were obtained from our numerical simulations on the dual Majorana monitored circuit. We extracted the correlation length exponents $\nu=1.75^{+0.12}_{-0.06}$, which is consistent with the previous study\cite{GYZhu2025SelfDual} of a transition captured by the same NLSM description but with a completely different microscopic origin unrelated to our decoding problem. 

Furthermore, we investigated the same decodability transition directly from the error correction perspective. We numerically computed the (post–optimal-decoding) logical error rates in a rotated square-lattice surface code subject to the same two-parameter coherent error model. The resulting decodability transition (with $\theta_1 \neq \theta_2$) corroborates the results from our numerical simulations of the dual Majorana monitored dynamics.

Our work points to several interesting directions for future research. First, we have shown that the hTC with $X$-type coherent error exhibits two distinct undecodable phases: one corresponding to the critical phase in the dual Majorana dynamics, and the other to an area-law phase. What is the difference between the decodability transitions into these two phases from the error-correction perspective? To address this question, it would be valuable to directly study the post-decoding logical error rates in the honeycomb-lattice toric code or surface code under the two-parameter error model.

Second, in the sTC decodability phase diagram [Fig.~\ref{fig:squarephase}(a)], the phase boundary between the decodable and undecodable area-law phases is expected to extend into the orange region (which is dominated by finite-size effects in our numerics). However, due to the limited system sizes accessible in our simulations, our data cannot reliably distinguish these two phases within this region. As discussed above, it is possible that the decodable phase with topological index $\cI=0$ includes the entire interval $0 < \theta_1 = \theta_2 < \pi/4$. Consequently, it is possible that the phase boundary between the two area-law phases extends all the way to the special point $\twotheta = (\pi/4, \pi/4)$. Larger-scale simulations, both in the dual Majorana dynamics and in the logical error rates of the rSC, would help clarify this scenario further.

Furthermore, the phase diagram in Fig.~\ref{fig:squarephase}(a) contains only area-law phases with topological indices $\cI = 0$ and $1$. It would be interesting to explore whether introducing additional spatial variations in the error angles, beyond our minimal two-parameter model, could enable area-law phases with $\cI \neq 0,1$ in the sTC decodability phase diagram. Even if such phases are undecodable, it may still be fruitful to understand how they differ from an error-correction perspective, both at finite system sizes and in the thermodynamical limit.

Last but not least, our work has focused on $X$- and $Z$-type coherent errors. It is natural to generalize to single-qubit coherent errors of the form $\prod_l e^{\ii (\alpha_l X_l + \beta_l Y_l + \gamma_l Z_l)}$, as well as to more general correlated coherent errors, such as $\prod_{\langle l, l' \rangle} e^{\ii \theta X_l X_{l'}}$. In these cases, the dual dynamics should be interacting \cite{BaoUnitaryErrors2024,BehrendsBeri2025}. Interestingly, on the sTC, the coherent error $\prod_l e^{\ii (\alpha_l X_l + \beta_l Y_l + \gamma_l Z_l)}$ respects the TR symmetry [Eq.~\eqref{eq:TR2}], whereas $\prod_{\langle l, l' \rangle} e^{\ii \theta Z_l Z_{l'}}$ does not. Ref.~\onlinecite{bhuiyan2025free} has extended the AZ symmetry classification framework to interacting monitored dynamics. It would be interesting to investigate how the symmetry classification affects the corresponding decoding problems and their dual interacting dynamics. Going beyond purely coherent error models, one can further generalize the investigation to include errors with both coherent and incoherent components. It would be valuable to understand how the symmetry classification influences the universal properties of decoding problems in this more general setting.

\section*{Acknowledgement}
We thank Yimu Bao, Benjamin B\'{e}ri, Asadullah Bhuiyan, Yuri Lensky, Simon Trebst, Sagar Vijay, and Guo-Yi Zhu for helpful discussions. We are grateful to the KITP, which is supported by the National Science Foundation under Grant No. NSF PHY-2309135, 
the KITP Program ``Learning the Fine Structure of Quantum Dynamics in Programmable Quantum Matter", the KITP Program ``Noise-robust Phases of Quantum Matter", and the KITP Conference ``Frontiers of Programmable Quantum Dynamics: Advances and Applications," where parts of this work were carried out. Y.Z. and C.-M.J. are supported by the Alfred P.
Sloan Foundation through a Sloan Research Fellowship.

{\it Note added.} - While we were finalizing this manuscript, we became aware of a related work\cite{VijayBaoYan2026} that partially overlaps with our discussion on the square-lattice toric code with uniform coherent error and the emergence of different symmetry classes on different lattices.

\appendix
\section{Mapping of error-corrupted toric codes to statistical models and Majorana monitored circuits}
In this appendix, we derive the mapping of the syndrome probability of error-corrupted toric code to the disordered classical Ising model [Eq. \eqref{eq:PtoZ}]. Once we have the partition function of the Ising model, we can apply the transfer-matrix method and the Jordan-Wigner transformation to further map it to a Majorana monitored circuit. In particular, we show that the so-obtained Majorana circuit consists of unitary gates and generalized measurements that satisfy the conditions for a positive operator-valued measure. 

\subsection{Mapping to disordered classical Ising model} 

\label{app:statmodel}
For simplicity, consider the toric code (on a square or honeycomb lattice) with appropriate boundary conditions, such that it admits only one logical qubit. In the following, we use the case of $X$-type coherent error as an example for the derivation of the disordered statistical model.

Consider the logical state $\ket{\overline{0}}$ and turn on the $X$-type coherent errors. The error corrupted state is
\begin{equation}
    \ket{\cE_X}=\prod_l e^{\ii\theta_lX_l}\ket{\overline{0}}.
\end{equation}
Upon the measurements of the $B_p$ stabilizers, the state collapses according to the syndromes $\{s_p=\pm1\}$ with Born rule probability
\begin{align}
{\cal P}_X(\{s_p\}) &= \bra{\cE_X}\prod_p\left(\frac{1+s_p B_p}{2}\right) \ket{\cE_X}\nonumber \\
&= \left\lVert \prod_p\left(\frac{1+s_p B_p}{2}\right) \ket{\cE_X}\right\rVert^2 \nonumber \\
& = \left(\prod_l\cos\theta_l\right)^2 \left\lVert \sum_E \prod_{l\in E}\ii \tan\theta_l X_l\ket{\overline{0}}\right\rVert^2 .
\end{align}

In the last equality, we expanded the product $\prod_l\left(1+\ii \tan\theta_l X_l\right)$ into strings of Pauli $X$ on a set of links $E$ (which we call ``error strings"). These error strings must be compatible with the syndrome $s_p$. Define the error variable
\begin{equation}
    \eta_l = \begin{cases}
        +1, &l\notin E \\
        -1, &l\in E.
    \end{cases}
\end{equation}
It satisfies
\begin{equation}
    \prod_{l\in p}\eta_l = s_p, \quad \forall p. \label{eq:syndromecomp}
\end{equation}
This condition states that the error strings are open strings on the dual lattice that end at plaquettes $p$ with $s_p=-1$. Given one error string $E$, we can find all other strings by deforming it with closed loops on the dual lattice. There are two kinds of loops, the contractible ones and the uncontractible ones. The former are generated by the stabilizers $A_v$, while the latter corresponds to a logical $X$ operator. 

Define $X_E \equiv \prod_{l\in E} X_l$. Let $E+E'$ denote the symmetric difference between $E,E'$ as sets of links, $E+E'\equiv E\cup E' - E\cap E'$. It follows that $X_{E+E'} = X_EX_{E'}$. Denote the contractible loops as $C$ and the uncontractible loop as $\overline{C}$. Suppose $E$ is compatible with $s_p$ [Eq. \eqref{eq:syndromecomp}]. Any other compatible error strings can be written as either $E+C$ or $E+C+\overline{C}$ for some $C$.  The logical operator $\overline{X}$ is given by $\overline{X}\equiv X_{\overline{C}}$. We can now proceed with the sum over error strings $E$,
\begin{widetext}
\begin{align}
    {\cal P}_X(\{s_p\}) &= \mathcal{N} \left\lVert \sum_C \left(\prod_{l\in {E+C}}\ii \tan\theta_l \right)X_{E+C}\ket{\overline{0}} +\sum_C \left(\prod_{l\in {E+C+\overline{C}}}\ii \tan\theta_l \right)\overline{X}X_{E+C} \ket{\overline{0}}\right\rVert^2 \nonumber\\
    &= \mathcal{N}\left\lVert w(E)\ket{\overline{0}} + w(E+\overline{C})\ket{\overline{1}} \right\rVert^2 \nonumber \\
    & =  \mathcal{N}(|w(E)|^2 + |w(E+\overline{C})|^2), \\
    {\rm with}~~~w(E) &\equiv \sum_C \prod_{l\in E+C}\ii\tan\theta_l, \quad \mathcal{N} \equiv \left(\prod_l\cos\theta_l\right)^2 .
\end{align}
\end{widetext}
In the second line above, we have used the stabilizer condition $X_C\ket{\overline{0}} =\ket{\overline{0}}, \forall C$. 

We observe that the central quantity in the Born rule probability is the weight $w(E)$, which consists of a sum over contractible loops on the dual lattice. The same loop summation can arise from the domain wall expansion of a classical Ising model. Indeed, let's put one Ising spin $\tau_v$ on each vertex $v$ and consider the Hamiltonian of the classical Ising model with random bond $\eta_l$,
\begin{equation}
    H = \sum_{l=\braket{v,v'}} \eta_lJ_l\tau_v\tau_{v'}.\label{eq:rbim}
\end{equation}
The corresponding partition function is, 
\begin{align}
    \cZ_{X,\{\eta_l\}} &= \sum_{\{\tau_v=\pm1\}} \exp\left(-\sum_{l=\braket{v,v'}}\eta_lJ_l\tau_v\tau_{v'}\right) \nonumber\\
    &= e^{-\sum_lJ_l}\sum_{\{\tau_v=\pm1\}}  \exp\left(\sum_{l=\braket{v,v'}}J_l(1-\eta_l\tau_v\tau_{v'})\right) \nonumber \\
    &=\cA e^{-\sum_lJ_l} \sum_C\prod_{l\in E+C} e^{2J_l}. \label{eq:app_parition}
\end{align}
In the last line, we have applied a domain wall expansion. $C$'s are domain walls, which are contractible loops on the dual lattice. $\tau_v\tau_{v'} = -1$ when $\braket{v,v'}\in C$. $\cA = 2$ if the system has a global $\mathbb{Z}_2$ symmetry and $\cA=1$ otherwise. By comparing the last line with $w(E)$ above, we identify
\begin{equation}
    e^{2J_l} = \ii\tan\theta_l.
\end{equation}
The syndrome probability and the partition function are then related by
\begin{equation}
    {\cal P}_X(\{s_p\}) \propto    |\cZ_{X,\{\eta_l\}}|^2 +|\cZ_{X,\{\overline{\eta}_l\}}|^2, \label{eq:syndromeborn}
\end{equation}
where $\overline{\eta}_l = -1$ when $l\in E+\overline{C}$. In the main text [Eq. \eqref{eq:PtoZ}], we have omitted the second term because the decodability transition is a bulk property, while the insertion of $\overline{C}$ depends on boundary conditions. Hence, the decodability transition is captured already by the singularity of the first term.

\subsection{Mapping to Majorana monitored circuit} \label{app:mapcirc}
In the following, we demonstrate the mapping between the disordered classical Ising models obtained from the error-corrupted toric codes and Majorana monitored circuits. 

For concreteness, we consider the disordered classical Ising model Eq. \eqref{eq:rbim} on the honeycomb lattice (Fig. \ref{fig:MajCircuit_hTCX} left panel) as an example. Rewrite the partition function using the transfer matrix. With time direction going vertically up, each horizontal bond becomes a two-spin $ZZ$ gate, 
\begin{equation}
    e^{-\eta_lJ_l Z_jZ_{j+1}} = e^{-\eta_l(\kappa_l+ \ii\frac{\pi}{4})Z_jZ_{j+1}}, \quad \kappa_l\equiv \frac{1}{2}\log\tan\theta_l.
\end{equation}
Each non-horizontal ($60^\circ$ and $120^\circ$) bond becomes a single-spin $X$ gate
\begin{equation}
    e^{\ii (\theta_l + \frac{\pi}{4}(1-\eta_l))X_j}.
\end{equation}
Now, apply the Jordan-Wigner transformation to transform the spins to Majorana fermion modes:
\begin{equation}
    X_j = \ii \psi_{2j-1}\psi_{2j}, \quad Z_jZ_{j+1} =\ii \psi_{2j}\psi_{2j+1} .
\end{equation}
The horizontal bond becomes the gate
\begin{equation}
    M_{\eta_l} = \frac{1}{\sqrt{2\cosh2\kappa_l}} e^{-\eta_l\kappa_l\ii \psi_{2j}\psi_{2j+1}}e^{\eta_l\frac{\pi}{4}\psi_{2j}\psi_{2j+1}} .
\end{equation}
The normalization factor ensures that $M_{\eta_l =\pm1}$ form a Kraus-operator set ${\cM}_l$ for a generalized measurement. That means the Kraus operators need to satisfy the condition for a positive operator-valued measure:
\begin{equation}
    \sum_{\eta_l = \pm1} M^\dag_{\eta_l}M_{\eta_l} = \mathds{1}.
\end{equation}
Physically, we can interpret $\{M_{\eta_l=\pm1}\}$ as the Kraus-operator set for a weak measurement of the local fermion parity $\ii \psi_{2j}\psi_{2j+1}$ followed by a swap of the two fermions. $\eta_l=\pm1$ is the measurement outcome. 

Each non-horizontal bond becomes the unitary gate
\begin{equation}
    U_{\eta_l} = e^{- (\theta_l + \frac{\pi}{4}(1-\eta_l))\psi_{2j-1}\psi_{2j}}.
\end{equation}
Here, $\eta_l=\pm1$ is a random parameter that takes each value with equal probability. 

In summary, the disordered classical Ising model partition function that appears in the syndrome probability is now mapped to a Majorana monitored circuit. Let's denote this circuit by $\cV_{\{\eta_l\}}$ for the quantum trajectory labeled by $\{\eta_l\}$ and the inital state by $\ket{\psi_0}$. The Born rule probability for the quantum trajectory
$\{\eta_l\}$ is
\begin{equation}
    \cP_{\rm circ}(\{\eta_l\}) = \bra{\psi_0}\cV^\dag_{\{\eta_l\}}\cV_{\{\eta_l\}}\ket{\psi_0}. \label{eq:circuitborn}
\end{equation}
An exact mapping of the partition function $\cZ_{X,\{\eta_l\}}$ to a circuit also involves a final state projection that corresponds to the boundary condition of the classical Ising model:
\begin{equation}
    \cZ_{X,\{\eta_l\}} = \bra{\psi_f}\cV_{\{\eta_l\}}\ket{\psi_0}.
\end{equation}
Comparing Eqs. \eqref{eq:syndromeborn} and \eqref{eq:circuitborn}, we see that the syndrome-sampling Born rule and Majorana-circuit Born rule are equal up to boundary conditions. For large systems, the quantum trajectory sampling in the monitored circuit then faithfully reproduces syndrome sampling. Since the universal properties of bulk phase transitions should not be affected by the boundary conditions, we use the Majorana circuit to map out the decodability phase diagram [Fig. \ref{fig:honeyphase}(a) and Fig. \ref{fig:squarephase}(a)]. To study code-geometry-specific quantities such as the logical error rate Eq. \eqref{eq:ler}, we use a different algorithm \cite{bravyi2018correcting} to sample the syndromes more accurately, which is reviewed in App. \ref{app:algosample}.

\section{Numerical simulation algorithms}
\subsection{Algorithm for Majorana monitored circuit}\label{app:algocircuit}
We briefly outline the algorithm for simulating the Majorana monitored circuits [Figs. \ref{fig:TwoParameter_hTCX}(b) and  \ref{fig:TwoParameter_sTC}(b)]. Since both the Kraus operators in the measurements [Eq. \eqref{eq:KrausOp}] and the unitary gates [Eq. \eqref{eq:RandUnitary}] are Gaussian, a Gaussian initial state stays Gaussian upon the circuit evolution \cite{Bravyi2005lagrangian}. A fermionic Gaussian state (FGS) on a set of Majorana fermion modes $\psi_j$ is completely determined by its covariance matrix
\begin{equation}
    \Gamma_{jk} = \frac{\ii}{2}\braket{[\psi_j, \psi_k]}.
\end{equation}
We choose a random FGS as the initial state for the monitored circuit. For the monitored circuit dual to the sTC, we also impose TR symmetry on the initial state for faster numerical convergence. 

To track the effect of a quantum gate on the FGS, it suffices to understand how it updates the covariance matrix. For a weak parity measurement of $\ii\psi_j\psi_k$ with measurement strength $\kappa>0$ and outcome $\eta = \pm1$, the Kraus operator is 
\begin{equation}
    M'_\eta = \frac{1}{\sqrt{2\cosh 2\kappa}}e^{\eta\kappa(\ii\psi_j\psi_k)}. \label{eq:weakmeasapp}
\end{equation}
This updates the covariance matrix as follows,
\begin{widetext}
\begin{align}
    \Gamma &\rightarrow \left(\mathds{1} -\frac{\cosh 2\kappa -1}{\cosh 2\kappa}\Pi^{(jk)}\right) (\Gamma + \delta\Gamma)\left(\mathds{1} -\frac{\cosh 2\kappa -1}{\cosh 2\kappa}\Pi^{(jk)}\right) - \eta\tanh2\kappa \Omega^{(jk)}, \nonumber \\
    \delta\Gamma& \equiv \frac{-\eta \tanh2\kappa}{1+\eta\tanh2\kappa\,\Gamma_{jk}}\Gamma \Omega^{(jk)}\Gamma, \label{eq:fgsM}
\end{align}
\end{widetext}
where $\Pi^{(jk)}$ is the projector onto the subspace spanned by $\{e_j, e_k\}$. $\Omega^{(jk)}$ acts as the symplectic matrix in this subspace, i.e.,
\begin{equation}
    \Omega = \begin{pmatrix}
        0 & -1\\
        1 & 0
    \end{pmatrix},
\end{equation}
and as a zero matrix outside this subspace. The Born-rule probability for the outcome $\eta = \pm1$ is
\begin{equation}
    \cP_\eta  = \frac{1+\eta \tanh2\kappa \Gamma_{jk}}{2}.
\end{equation}

A two-fermion unitary operator
\begin{equation}
    U = e^{-\theta\psi_j\psi_k} \label{eq:unitaryapp}
\end{equation}
updates the covariance matrix by a simple rotation in the $(jk)$ subspace:
\begin{align}
\Gamma \to R^{(jk)}\Gamma (R^{(jk)})^{\mathsf T},\quad
R = \begin{pmatrix}
  \cos2\theta & -\sin2\theta \\
  \sin2\theta & \cos 2\theta
\end{pmatrix}. \label{eq:fgsU}
\end{align}
Note that the generalized measurement operator Eq. \eqref{eq:KrausOp} in the main text can be obtained by applying the weak measurement Eq. \eqref{eq:weakmeasapp}  followed by a unitary rotation Eq. \eqref{eq:unitaryapp} with $\theta=\pm\pi/4$. 

For a subsystem $A$, the subsystem entanglement entropy (EE) $S_A$ in a Gaussian state with covariance matrix $\Gamma$ is given by
\begin{equation}\label{eq:f}
    S_A =-\tr\left(\frac{\mathds{1}+\ii\Gamma_A}{2}\log\left(\frac{\mathds{1}+\ii\Gamma_A}{2}\right)\right).
\end{equation}
Here, $\Gamma_A$ is the covariance matrix $\Gamma$ restricted to the fermion modes within $A$. The calculation of the mutual information $I_{A,B} = S_A+S_B - S_{AB}$ between two subsystems $A$ and $B$ follows from the expression above.

We simulate the Majorana monitored circuits dual to sTC and hTC on Majorana chains with periodic boundary conditions. The number of Majorana fermions $L$ is always a multiple of $16$. For a size $L$ system, the number of timesteps is fixed to $T=L/2$ for hTC and $T=L/4$ for sTC. Each timestep consists of four layers of unitary or measurement gates, as plotted in Figs. \ref{fig:MajCircuit_hTCX} and \ref{fig:MajCircuit_sTC}. This choice of circuit depth keeps the aspect ratio of the toric codes of order unity. 

To speed up the sampling of EE and MI in our monitored circuit simulation, we perform a ``sliding average" in addition to the direct sampling over quantum trajectories. For EE and MI, we first divide the periodic fermion chain into two or four equal-size connected subsystems and also fix the relevant subsystems $(A,B)$ for MI. This gives one sample of EE or MI per quantum trajectory. We then slide the boundaries between the subsystems by steps of 4 fermion sites, yielding one sample for each unique division and subsystem-choice resulted from such moves. For example, for $L=64$, we obtain 8 unique samples for both EE and MI per quantum trajectory.

\subsection{Algorithm for syndrome sampling and maximal likelihood decoder for rotated surface code}\label{app:algosample}

\begin{figure}[ht!]
    \centering
    \includegraphics[width=.9\linewidth]{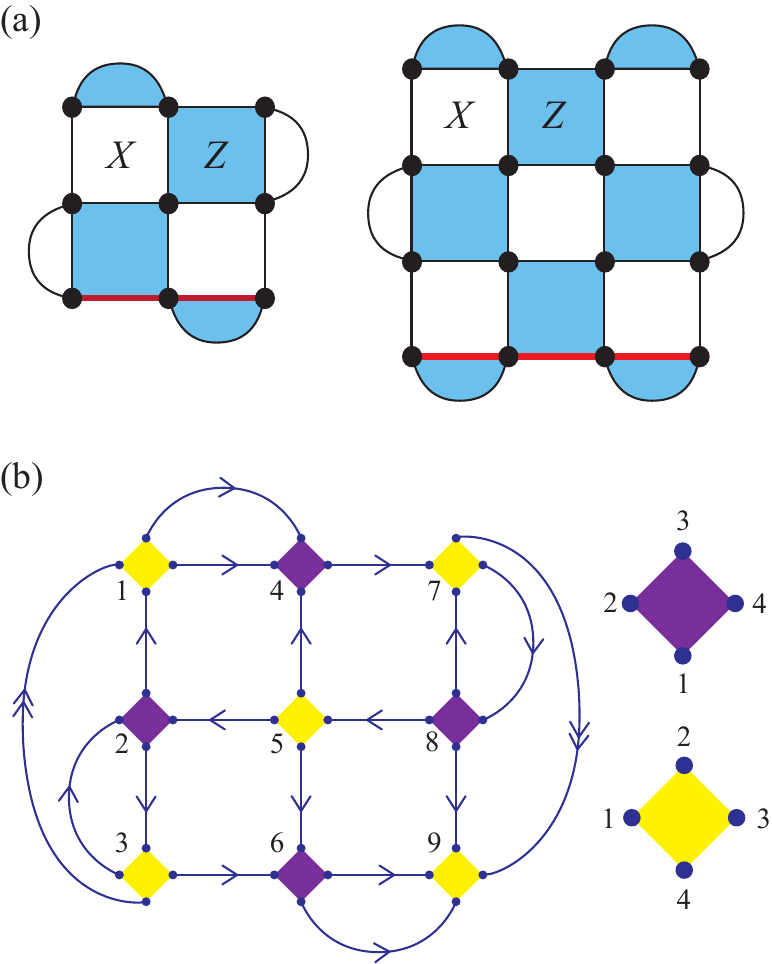}
    \caption{(a) Rotated surface code (rSC) geometries for odd (left panel) and even (right panel) code distances. Each black dot is a qubit, totaling $d^2$ qubits for a distance-$d$ code. Each shaded square or cap indicates a $Z$ stabilizer, and the unshaded shapes indicate $X$ stabilizers. A logical $X$ operator acts on a horizontal row of qubits (red line). (b) Mapping surface code to a 2+1D Majorana circuit: we show the geometry with $d=3$ as an example. Each qubit (purple or yellow diamond) is replaced by four Majorana modes (blue corners of the diamond). The Majorana modes in each diamond are ordered according to $c_1,\dots, c_4$ as shown in the right panel. The initial state of this Majorana circuit is dimerized according to the arrows in the left panel. An arrow pointing from $c_j$ to $c_k$ indicates a dimer selected by the projection $\frac{1+ic_jc_k}{2}$. 
    For larger systems, the single arrows are repeated in a translationally invariant manner. The double arrows depend on the initial logical state of the surface code. Here, the choice of double arrows corresponds to the logical state $\ket{\overline{0}}$. The Gaussian gates in Eq. \eqref{eq:rscmajgate} are applied according to the number ordering in the left panel.}
    \label{fig:rSC}
\end{figure}

The rotated surface code (rSC) geometry is shown in Fig. \ref{fig:rSC}(a). A distance-$d$ code has $d^2$ physical qubits. We briefly review the sampling algorithm for the syndromes generated by $X$-type coherent errors and refer the readers to the original paper \cite{bravyi2018correcting} for details. 

The first step is to map the qubits to fermions. This is done by embedding each qubit into four Majorana modes $c_1,c_2,c_3,c_4$. The qubit Hilbert space can be recovered by the projection
\begin{equation}
    \frac{1+S}{2} ~~{\rm with}~~ S = -c_1c_2c_3c_4.
\end{equation}
The Pauli operators on the qubit are given by
\begin{equation}
    X = \ii c_2c_3 = \ii c_1c_4 S,\quad Z = \ii c_1c_2 = \ii c_3c_4S.
\end{equation}
The syndrome sampling of the rSC is then mapped to a 2+1D Majorana Gaussian circuit (which is not a monitored circuit with non-postselected measurements). The initial state $\ket{\phi_0}$ is a Gaussian state with dimerization pattern shown in Fig. \ref{fig:rSC}(b). Then, a Gaussian gate is applied to each qubit site according to the order shown in Fig. \ref{fig:rSC}(b). Consider applying the $l$-th gate $G_l$ to the Majorana state $\ket{\phi_{l-1}}$. The gate $G_l$ consists of a unitary rotation followed by two fermion parity projections
on the four Majorana modes on the $l$-th site: 
\begin{align}
    G_l = \frac{1+\eta_l \ii c_1 c_2}{2} \frac{1+\eta_l \ii c_3 c_4}{2} \exp(-\theta_lc_2c_3). \label{eq:rscmajgate}
\end{align}
Here, $\eta_l$ is the random bond variables that capture the syndromes via $s_p = \prod_{l\in p} \eta_l$. They are located on the links in the sTC, which corresponds to the lattice sites in the rSC. The syndromes $s_p$ are located in the shaded squares in Fig. \ref{fig:rSC}(a).  We remark that this gate in Eq. \ref{eq:rscmajgate} cannot be interpreted as a measurement.
That is because the two realizations of $G_l$ with $\eta_l=\pm 1$ do not combine into a Kraus-operator set that obeys the condition for a positive operator-valued measure. For this reason, this 2+1D Majorana circuit for the syndrome sampling is not a monitored circuit with genuine (non-postselected) measurements. Nevertheless, this 2+1D Majorana circuit can be computed efficiently due to its Gaussian nature.

The application of the gate $G_l$ updates the covariance matrix of the Gaussian state $\Gamma_{l-1}\to \Gamma_l$. The update rule upon unitary rotations is given by Eq. \eqref{eq:fgsU}. The update rule upon the projection $\frac{1+\eta \ii c_jc_k}{2}$ is obtained by taking $\kappa\to\infty$ in Eq. \eqref{eq:fgsM}. It takes the simpler form,
\begin{align}
    &\quad\Gamma\to \nonumber \\
    &(\mathds{1}-\Pi^{(jk)})\left(\Gamma - \frac{\eta}{1+\eta\Gamma_{jk}}\Gamma\Omega^{(jk)}\Gamma\right)(\mathds{1}-\Pi^{(jk)}) - \eta\Omega^{(jk)}.
\end{align}

The conditional probability for the random bond variable $\eta_l$ given $\eta_1,\dots,\eta_{l-1}$ is
\begin{equation}
    \cP(\eta_l|\eta_1,\dots,\eta_{l-1}) = \mathcal{N}_l \left\lVert G_l\ket{\phi_{l-1}}\right\rVert^2,
\end{equation}
where $\mathcal{N}_l = 2$ for the last qubit $l=d^2$ and $\mathcal{N}_l = 1$ otherwise. Since $\ket{\phi_{l-1}}$ is a Gaussian state, we can calculate the RHS with the covariance matrix $\Gamma_{l-1}$ and Wick's theorem. At each time step, we apply a gate $G_l$, sample $\eta_l$ with this conditional probability, and update the covariance matrix $\Gamma_{l-1} \rightarrow \Gamma_l$ accordingly. After applying all $d^2$ gates, we finish the sampling of one random bond realization $\{\eta_l\}$. We can obtain a total probability $\cP(\{\eta_l\})$ for the random bond configuration $\{\eta_l\}$ by the product of the conditional probabilities above.

One can show that this total probability satisfies $\cP(\{\eta_l\}) \propto |\cZ_{X,\{\eta_l\}}|^2$ (as defined in Eqs. \eqref{eq:app_parition}). Hence, it is proportional to the probability shared by all the random-bond configurations in the same error class as $\{\eta_l\}$. To obtain the probability for the opposite error class $\{\overline{\eta}_l\}$, we insert a logical error $\overline{X}$ by flipping all $\eta_l\to -\eta_l$ for the lowest rows of the qubits in the rSC. This gives another sampling probability $\cP(\{\overline{\eta}_l\})$. Suppose the decoder chooses $\{\eta_l\}$ as the error class for the correction Pauli string $\cC_s$. The logical error rate can be calculated,
\begin{equation}
    P_L = \frac{\cP(\{\overline{\eta}_l\})}{\cP(\{\eta_l\}+\cP(\{\overline{\eta}_l\}}.
\end{equation}
The maximal likelihood decoder is one that always chooses the error class for the correction string that minimizes $P_L$.

For memory efficiency, it is not necessary to store the entire covariance matrix $\Gamma_l$ at any given step. Since initial states are dimerized, the covariance matrix is formed by blocks of $\Omega$. Moreover, any fermions dimerized by gate $G_l$ will remain in that state and no longer be active in subsequent calculations. Hence, at any step $l$, we only need to store $\Gamma$ restricted to a subspace involving fermions which either appear in the gate $G_l$ or are not dimerized. This turns out to be always a subspace of size about $d$, which is effectively a 1D subsystem of the fermions in the 2D lattice.

\section{Algorithm for scaling collapse}
\label{app:algo_ScalingCollapse}

Consider a statistical physics system with a tuning parameter $\theta$. Suppose a continuous phase transition happens at a critical point $\theta=\theta_c$. Near the critical point, an observable $y$ generally follows the scaling ansatz,
\begin{equation}
    y(\theta,L) = L^{-\Delta}f\!\left((\theta-\theta_c)L^{1/\nu}\right), \label{eq:fss}
\end{equation}
where $L$ is the system size, $\nu$,$\Delta$ are critical exponents and $f$ is an unknown function. In this paper, $y$ is the quarter-subsystem mutual information $I_{A,B}$ for the Majorana monitored circuit. In this case, $\Delta=0$ \cite{LiConformal2021}.

Given data points $(\theta_i,y_i, L)$ and an estimate of parameters $(\theta_c, \nu)$, we can plot the observable $y_i$ against the scaled coordinates,
\begin{equation}
    x_i = (\theta_i-\theta_c)L_i^{1/\nu},
\end{equation}
For each system size $L$, this produces a curve $y_L(x)$. If the estimated parameters are accurate, the ansatz Eq. \eqref{eq:fss} (with $\Delta=0$) implies $y_L(x)= f(x), \forall L$, i.e., all curves $y_L$ collapse onto a single function. The quality of data collapse is quantified by a weighted loss function, which we now define.

Consider a data point $(x_0,y_0)$ obtained at a given system size $L$, with standard error $\delta y_0$. To test the quality of scaling collapse at this point, we compare it to a local estimate of $f(x)$ constructed from data points at all other systems sizes $L'\neq L$. For each system size $L'\neq L$, if $x_0$ lies within the range of that dataset, we identify two data points $(x_1,y_1)$ and $(x_2,y_2)$ such that $x_1$ and $x_2$ are the nearest available scaled coordinates immediately below and above $x_0$. Repeating this procedure for all system sizes $L'\neq L$ yields a collection of reference data points $(x_\alpha, y_\alpha)$, with $x_\alpha$ near $x_0$, each associated with a standard error $\delta y_\alpha$. These reference points are then used to perform a weighted least-squares fit to a local linear function
\begin{equation}
    Y(x) = m x + b,
\end{equation}
with weights $w_\alpha = 1/(\delta y_\alpha)^2$. This yields a prediction for the observable $Y_0 \equiv Y(x_0)$ with uncertainty $\delta Y_0$ obtained from the covariance matrix of the fit.

The contribution of the data point $(x_0,y_0)$ to the weighted loss function\cite{Houdayer2004lowtemp,pyfssa2015quality} is defined as
\begin{equation}
    \frac{(y_0 - Y_0)^2}{(\delta y_0)^2 + (\delta Y_0)^2}.
\end{equation}
The total loss function is obtained by averaging this quantity over all valid data points. The optimal values of $(\theta_c,\nu)$ are determined by minimizing this loss function. A good scaling collapse produces a loss function of order unity, indicating that the curves $y_L(x)$ at different system sizes $L$ agree within their statistical uncertainties.

To estimate uncertainties in the fitted parameters $(\theta_c,\nu)$ and assess finite-size effects, we perform pairwise scaling collapses \cite{jian2023measurement}. For every pair of system sizes $(L_1,L_2)$, we consider the subset of data points obtained at these two sizes and repeat the fitting procedure described above, obtaining estimates
\begin{equation}
    \theta_c(L_1,L_2), \quad \nu(L_1,L_2).
\end{equation}
We quote error bars based on the extrema of these pairwise fitted values. An example of this pairwise fitting is shown in Fig. \ref{fig:fss}(e) and (f).

\section{Scaling collpase data} \label{app:datafss}

\begin{figure*}
    \centering
    \includegraphics[width=0.9\linewidth]{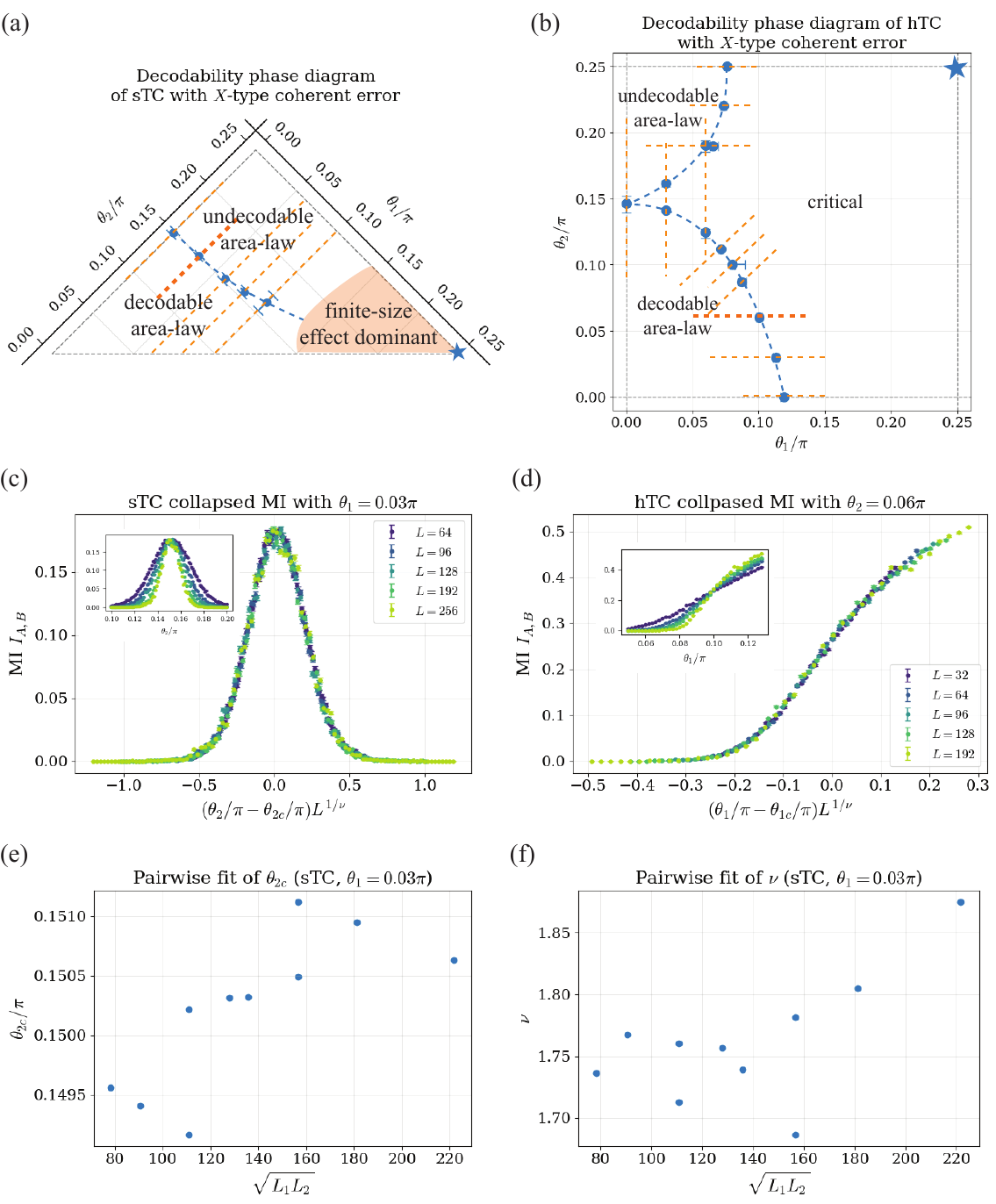}
    \caption{Data for the scaling collapse of mutual information (MI) scans. (a)(b) Decodability phase diagram of the sTC and the hTC models with the MI scans indicated by orange dashed lines. The phase boundaries are determined by the scaling collapse of these MI scans. The MI data are averaged over 5000-10000 samples. (c) Scaling collapse of a representative MI scan in the sTC phase diagram with $\theta_1 = 0.03\pi$ (thick orange dashed line in (a)). The fitted parameters are $\theta_{2c}/\pi = 0.1504^{+0.0007}_{-0.0012}, \nu=1.75^{+0.12}_{-0.06}$. Inset: MI scan before collapse. (d) Scaling collapse of a representative MI scan in the hTC phase diagram with $\theta_2=0.06\pi$ (thick orange dashed line in (b)). The fitted parameters are $\theta_{1c}/\pi = 0.1000^{+0.0016}_{-0.0019}, \nu=2.28^{+0.17}_{-0.23}$. Inset: MI scan before collapse. (e)(f) Uncertainty analysis of the fitted paramters $\theta_{2c}, \nu$ for the scaling collpase in (c). Each parameter is refitted using a subset of the data with only two system sizes $L_1, L_2$. All such pairwise estimates of the parameters are plotted. The extrema of these estimates are taken as the asymmetric error bars of the parameters.}
    \label{fig:fss}
\end{figure*}

The 1D scans on the $(\theta_1, \theta_2)$ parameter space used to determine the phase boundaries for the sTC and hTC models are shown in Figs. \ref{fig:fss}(a) and (b).

Representative scaling collapses for both models are shown in Figs. \ref{fig:fss}(c) and (d). 

The pairwise estimates of the critical angle $\theta_{2c}$ and the critical exponent $\nu$ for the scaling collapse of the data in Fig. \ref{fig:fss}(c) are shown in Figs.\ref{fig:fss}(e) and (f).

\section{Evidence for strong finite-size effect in the two-parameter sTC model}\label{app:finitesize}
\begin{figure*}[t!]
    \centering
    \includegraphics[width=0.9\linewidth]{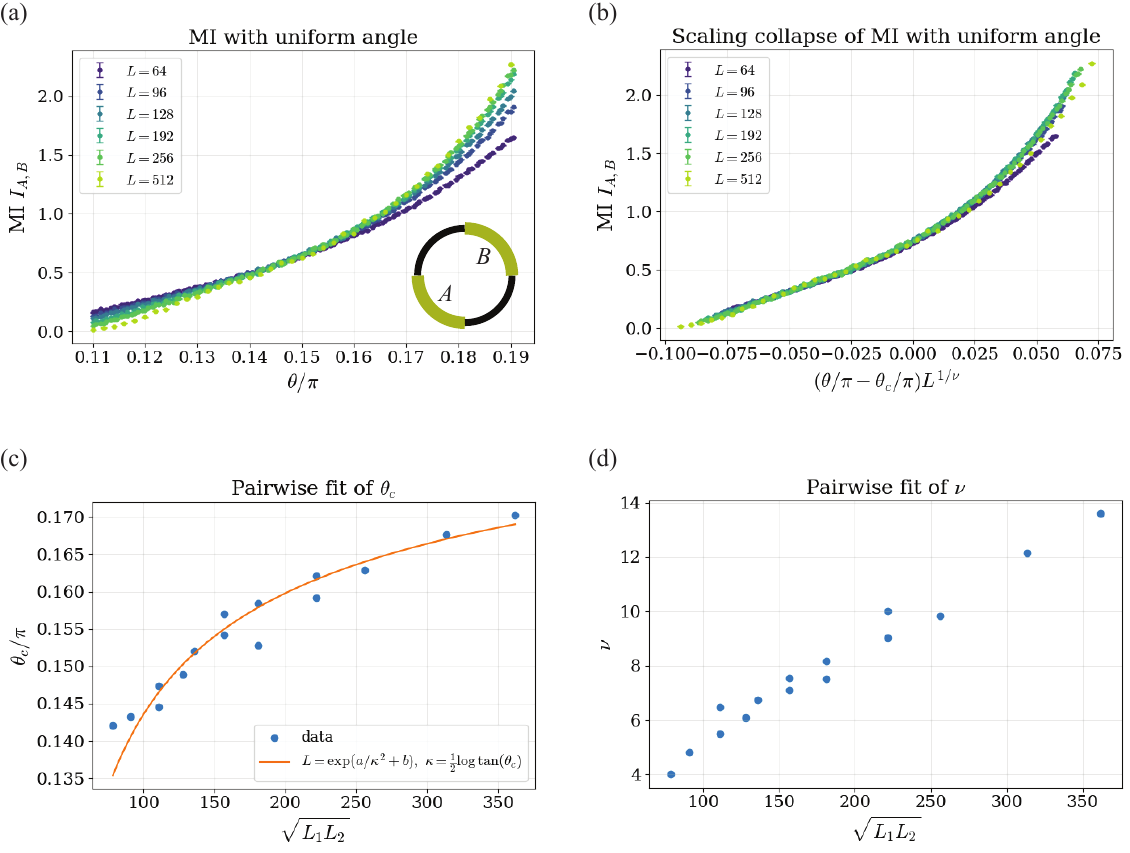}
    \caption{Finite-size scaling analysis of mutual information (MI) for the final state in the monitored circuit dual to sTC with a uniform error angle $\theta_1=\theta_2= \theta$ (a) MI is averaged over 5000 samples. MI curves of different system sizes appear to cross near $\theta= 0.15\pi$. (b) Data collapse assuming the crossing of MI corresponds to a continuous phase transition at a critical $\theta_c$. Fitted parameters $\theta_c = 0.155\pi, \nu=8.6$. The data collapse quality is visibly poor. (c)(d) Drift of parameters for scaling collapse on subsets of data with two system sizes $L_1, L_2$. (c) Drift of $\theta_c$ is consistent with the slow RG flow (Eq. \eqref{eq:RG_ClassD}) away from the unitary point $\theta=\pi/4$. The fitting curve is derived from the RG flow. (d) $\nu$ drifts to larger and larger values for increasing $L$, suggesting the breakdown of the assumption of a continuous phase transition. }
    \label{fig:drift}
\end{figure*}
In this appendix, we present evidence that for the two-parameter sTC model, in the parameter region near $(\theta_1,\theta_2) = (\pi/4, \pi/4)$ [see the orange region in Fig. \ref{fig:squarephase}(a)], the numerics are dominated by strong finite-size effects. This potentially explains the tension between the previously observed critical-like phase \cite{Venn_2023}, and the fact that class-D Majorana monitored circuits in 1+1D have no stable critical phase under RG.

\subsection{Drifting of fitted parameters of scaling collapse}

In the sTC under the uniform error model $\theta_1=\theta_2=\theta$, consider a scan of MI into the region dominated by finite-size effects $\theta\gtrsim0.15\pi$ [Fig. \ref{fig:drift}(a)]. Like previous cases, this MI refers to the MI $I_{A,B}$ between two antipodal quarter-system-size regions $A$ and $B$. If there is a true critical phase and an area-law-to-critical phase transition, one should be able to collapse the MI data onto a single scaling function.

A scaling collapse is shown in Fig.\ref{fig:drift}(b). The ansatz for scaling collapse takes the standard form:
\begin{equation}
    I_{A,B}(L,\theta) = f\left((\theta - \theta_{c})L^{1/\nu}\right), \label{eq:FSSIAB}
\end{equation}
where the critical angle $\theta_{c}$ and the critical exponent $\nu$ are fitting parameters. The quality of the scaling collapse is visibly poor, suggesting a possible breakdown of the ansatz.

For genuine continuous transitions, the fitted parameters $\theta_{c},\nu$ are expected to converge as the system size $L$ increases. For the present case, the fitted parameters $\theta_{c},\nu$ in the pairwise fitting scheme are shown in Figs. \ref{fig:drift}(c) and (d). 
It appears that $\nu$ drifts to larger and larger values as the system size $L$ increases. 
This is another sign that this system under investigation might not follow the general ansatz Eq. \eqref{eq:FSSIAB} of scaling collapse. 

The drifting of $\theta_c$ in Fig.\ref{fig:drift}(d) potentially provides more clues. We first attempt to fit $\theta_c$ to the following form, which is appropriate for second-order transitions,
\begin{equation}
    \theta_c(L) = \theta_{c}(\infty) +a L^{-b},
\end{equation}
where $a, b$ are fitting parameters. Physically, $b$ should be related to the correlation length exponent $\nu$ by $b=-1/\nu$ and $\theta_{c}(\infty)$ would be the critical value of $\theta$ in the thermodynamical limit. In this case, the best fit value is $\theta_{c}  = 5.1 \pi$ with 68\% confidence interval $(0.2\pi,8.1\pi)$. However, the symmetry of the phase diagram Fig. \ref{fig:squarephase}(a) constrains $\theta_c\leq0.25\pi$ (if there is an area-law-to-critical transition). The fact that the large best fit value $\theta_{c}=5.1\pi>0.25\pi$ suggests that the critical angle might not exist. 

Alternatively, we fit the drifting critical point $\theta_c(L)$ to an ansatz inspired by the beta function Eq. \eqref{eq:RG_ClassD}:
\begin{equation}
    L = e^{a/\kappa^2+b},~~~ \kappa = \frac{1}{2}\log\tan\theta_{c}
\end{equation}
with $a>0$. Here, the idea is that the physical system can appear critical if the correlation length is comparable to the system size $L$. Here, we assume $\theta=0.25\pi$ is the weak-coupling limit $g\rightarrow 0$ of Eq. \eqref{eq:RG_ClassD}. 
At the phase boundary, we identify $L$ as the correlation length $\xi$ (predicted by RG), since the drifting phase boundary $\theta_{c}(L)$ should appear when the correlation length $\xi$ grows beyond the system size $L$ as $\theta$ approaches $0.25\pi$ and the system appears scaleless. $\kappa$ is the strength of weak measurements (as defined in Eq. \eqref{eq:KrausOp}), which we conjecture to be proportional to the bare coupling constant $g$ in the effective field theory Eq. \eqref{eq:NLSMsTC}. The quality of fitting is reasonable [Fig. \ref{fig:drift}(c)], which also suggests the absence of the critical phase and the area-law-to-critical phase transition.  However, we caution that the data in Fig. \ref{fig:drift}(c) are obtained from the pairwise fitting scheme involving two system sizes $L_1,L_2$. We've used the crossing point of the two curves as a proxy for the drifting phase boundary $\theta_c(L=\sqrt{L_1L_2})$ associated with a single system size $L$.

\subsection{Behavior of two-point correlation}
\begin{figure*}[t!]
    \centering
    \includegraphics[width=\linewidth]{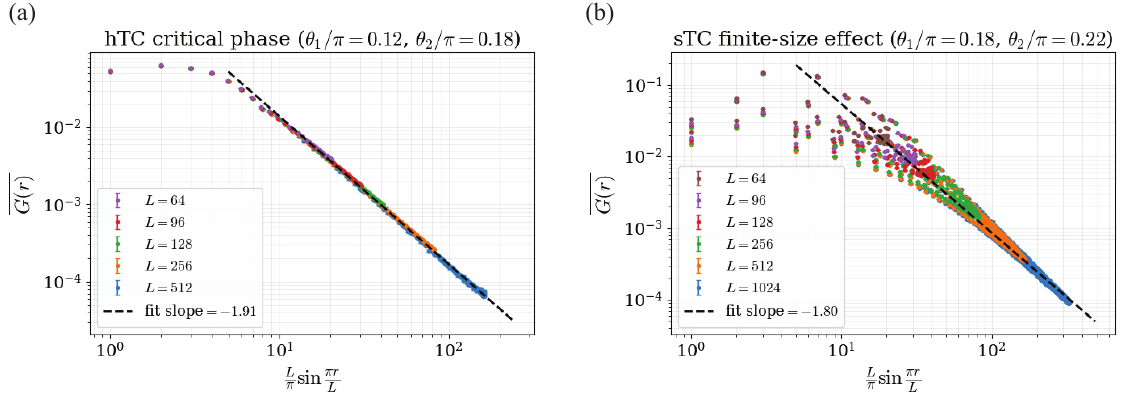}
    \caption{Comparison of the average second moment of the two-point correlation functions $\overline{G(r)}$ between (a) the critical phase of the hTC model with $X$-type coherent error and (b) the ``finite-size-effect-dominant" regime of the sTC model.
    The horizontal axis is the chord distance $d(r)$. (a) hTC model in the critical phase. A fitting to the power-law behavior Eq. \eqref{eq:G_powerlaw} yields $\Delta\approx 1.91$ (see black dashed line). 
     This power law correlation is a signature of conformal symmetry in the critical phase. 
    (b) sTC model in the region dominated by finite-size effects. $\overline{G(r)}$ does not fit well with a power-law behavior (black dashed line)
    The data for $L\leq256$ are averaged over 50000-100000 samples. The data for $L=512,1024$ are averaged over 1000-20000 samples. }
    \label{fig:twopointcorr}
\end{figure*}
Another clue for the strong finite-size effect in the orange region of the phase diagram Fig. \ref{fig:honeyphase}(a) comes from the average second moment of the two-point correlation function in the final state of the monitored circuit:
\begin{equation}
    \overline{G(r)} = \overline{|\braket{\ii \psi_n\psi_{n+r}}|^2},
\end{equation}
where the overline denotes the average over quantum trajectories. In a true critical phase, $\overline{G(r)} $ is expected to follow a power-law behavior on large systems
\begin{equation}
    \overline{G(r)} \propto \frac{1}{r^{\Delta}},
\end{equation}
with an exponent $\Delta$. For the critical phase in the class-DIII non-interacting Majorana monitored circuit, it was found to be $\Delta=2$ previously (modulo further logarithmic corrections) \cite{JianRTN2022,nahum2020entanglement}.

For a periodic chain of length $L$, conformal symmetry in the critical phase modifies the power law in $r$ into a power law in the chord distance,
\begin{equation}
    \overline{G(r)} \propto \frac{1}{d(r)^{\Delta}}, \quad d(r) = \frac{L}{\pi}\sin\frac{\pi r}{L}.
    \label{eq:G_powerlaw}
\end{equation}
We confirm this power-law behavior in the critical phase of the two-parameter hTC model with $X$-type coherent error [Fig. \ref{fig:twopointcorr}(a)]. We also plotted $\overline{G(r)}$ versus $d(r)$ for the two-parameter sTC model at one point in the finite-size dominant region [Fig. \ref{fig:twopointcorr}(b)]. The behavior does not fit well with a power-law behavior that would indicate a critical phase.

\bibliography{ref}

\end{document}